\documentclass{emulateapj}
\usepackage{txfonts}
\usepackage{graphicx}
\usepackage{subfigure}
\usepackage{multirow}
\usepackage{verbatim}
\usepackage{url}

\def\lsim{\ensuremath{\sim\kern-1em\raise 0.65ex\hbox{$<$}}}
\def\rsim{\ensuremath{\sim\kern-1em\raise 0.65ex\hbox{$>$}}}

\shorttitle{The First Astronomical Survey at $80^{\circ}$N}
\shortauthors{N.M. Law et al.}

\begin{document}

\title{Exoplanets from the Arctic: The First Wide-Field Survey at $80^{\circ}$N}

\author{Nicholas M. Law\altaffilmark{1}, Raymond Carlberg\altaffilmark{2}, Pegah Salbi\altaffilmark{2}, Wai-Hin Wayne Ngan\altaffilmark{2}, Aida Ahmadi\altaffilmark{3}, Eric Steinbring\altaffilmark{4}, Richard Murowinski\altaffilmark{4}, Suresh Sivanandam\altaffilmark{1}, Wolfgang Kerzendorf\altaffilmark{2}}
\altaffiltext{1}{\textit{law@di.utoronto.ca}; Dunlap Fellow, Dunlap Institute for Astronomy and Astrophysics, University of Toronto, 50 St. George Street, Toronto, Ontario M5S 3H4, Canada}
\altaffiltext{2}{Department of Astronomy and Astrophysics, University of Toronto, 50 St. George Street, Toronto, Ontario M5S 3H4, Canada}
\altaffiltext{3}{University of Calgary, 2500 University Dr. NW, Calgary, Alberta T2N 1N4, Canada}
\altaffiltext{4}{National Science Infrastructure, National Research Council Canada, Victoria, British Columbia, V9E 2E7, Canada}

\begin{abstract}
Located within 10 degrees of the North Pole, northern Ellesmere Island offers continuous darkness in the winter months. This capability can greatly enhance the detection efficiency of planetary transit surveys and other time domain astronomy programs. We deployed two wide-field cameras at $80^{\circ}$N, near Eureka, Nunavut, for a 152-hour observing campaign in February 2012. The 16-megapixel-camera systems were based on commercial f/1.2 lenses with 70mm and 42mm apertures, and they continuously imaged 504 and 1,295 square degrees respectively. In total, the cameras took over 44,000 images and produced better-than-1\% precision light curves for approximately 10,000 stars. We describe a new high-speed astrometric and photometric data reduction pipeline designed for the systems, test several methods for the precision flat-fielding of images from very-wide-angle cameras, and evaluate the cameras' image qualities. We achieved a scintillation-limited photometric precision of 1-2\% in each 10s exposure. Binning the short exposures into 10-min chunks provided a photometric stability of 2-3 millimagnitudes, sufficient for the detection of transiting exoplanets around the bright stars targeted by our survey. We estimate that the cameras, when operated over the full arctic winter, will be capable of discovering several transiting exoplanets around bright \mbox{($\rm m_V$ $<$ 9.5)} stars.
\end{abstract}


\maketitle

\section{Introduction}

The diurnal cycle is a significant impediment to ground-based transit surveys, limiting their detection efficiency for planets with orbital periods of tens of days or more \citep{Brown2003}. Consequently, exoplanet transit surveys rely on long-term observing campaigns and/or the use of multiple observing sites to reach an acceptable probability of detecting multiple transits. In contrast, circumpolar locations provide continuous darkness in the winter months, potentially allowing single-instrument ground-based surveys to reach longer period planets with high detection efficiencies. 

The detections of transiting planets around very bright stars (e.g. \citealt{Charbonneau2000, Henry2000, Sato2005, Bouchy2005, Christian2006, Bakos2007, Pal2010, Winn2011, Howell2012}), and the relative ease of characterization of those planets (e.g. \citealt{Agol2010, Collier2010, Demory2011, vonBraun2011, Stevenson2011, Majeau2012}), have demonstrated that very-small-aperture telescopes can discover exciting exoplanets. Here, we describe two circumpolar very-wide-field, small-aperture cameras designed to search for transiting exoplanets around 5,000 stars brighter than $\rm{m_V}$=9.5.

The advantages of circumpolar astronomy have already motivated several site characterization and observational projects based on the high Antarctic glacial plateau. The Gattini cameras have measured the optical sky brightness, cloud cover and auroral emission at both Dome A and Dome C \citep{Moore2006,Moore2008,Moore2010}, while small, robotically controlled, optical and infrared telescopes are in the testing and operation phases \citep{Rico2010, Daban2010}. The CSTAR project \citep{Wang2011} demonstrated long-term photometry on 10,000 stars in a 23 deg$^2$ region centered on the South Celestial Pole.

\begin{figure}
  \centering
  \resizebox{1.0\columnwidth}{!}
   {
	\includegraphics{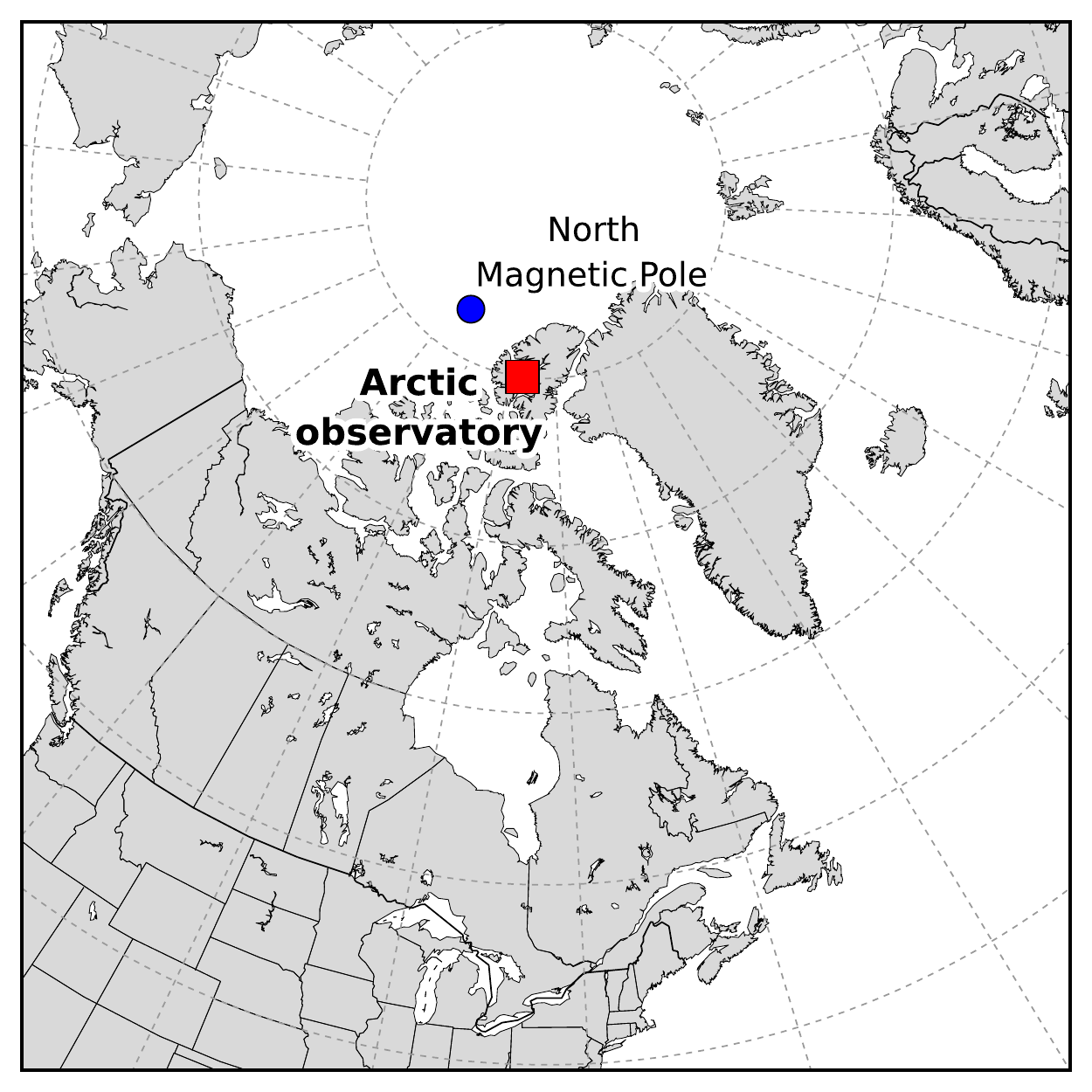}
   }
   \caption{The High Canadian Arctic location of the test system (80$^\circ$00'N, 86$^\circ$20'W) at the Ridge Lab near Eureka, Nunavut. The 2012 position of the North Magnetic Pole is also marked.}

   \label{fig:map}
\end{figure}

In this paper we present the first astronomical survey conducted from Ellesmere Island, located in the High Canadian Arctic at 80$^{\circ}$N in the territory of Nunavut (Figure \ref{fig:map}). Robotic site-testing systems deployed at nearby coastal mountain sites found a high clear-sky fraction, low median wind speed, and the possibility of excellent seeing \citep{Steinbring2010}. More intensive site testing at a lower-elevation manned station, the Polar Environment Atmospheric Research Laboratory (PEARL), showed photometric conditions for approximately half of the continuously dark polar night, and continuous periods of hundreds of hours of clear, dark skies. Furthermore, sufficiently clear conditions for spectroscopy or differential photometry occur for up to 86\% of the winter \citep{Steinbring2012}. Seeing measurements at the same location show the presence of a weak turbulence layer near the ground \citep{Hickson2010}, potentially providing excellent observing conditions.

The excellent conditions at the Ridge Lab site and their implications for exoplanet detection efficiencies, along with the existing infrastructure and relative ease of accessibility, encouraged us to design astronomical survey instruments for the site. In this paper we describe the design, operation and first results from two very-wide-field exoplanet-search cameras, the AWCam (Arctic Wide-Field Camera) prototypes, which we deployed to the Ridge Lab in February 2012. Although lasting only a week, and so within the regime for which a mid-latitude observatory could achieve similar efficiencies, this initial campaign was intended as a test of the survey capabilities achievable under real observing conditions at the site, as a prelude to longer operations in later years. 

The AWCam cameras continuously observed a several-hundred-square-degree field around the North Celestial Pole, using short exposures to avoid the need for a tracking mount. Like several lower-latitude transiting exoplanet surveys such as SuperWASP \citep{Pollacco2006}, HAT \citep{Bakos2004}, and KELT \citep{Pepper2007, Pepper2012}, the systems are based on commercial camera lenses. The cameras are primarily designed to search for transiting planets around bright ($\rm{m_V<10}$) stars, but the wide fields also provide continuous photometric coverage of tens of thousands of fainter stars, along with atmospheric transmission, cloud structure and auroral measurements. The initial survey demonstrated multi-color photometry on $\approx$70,000 stars, showed that the images can be precisely astrometrically and photometrically calibrated, and demonstrated that the systems can achieve millimagnitude-level photometry.

The paper is divided as follows. In section \ref{sec:site} we evaluate the expected transit detection efficiency at the Ellesmere Island site. In section \ref{sec:design} we describe the system design and construction, including details on the thermal protection systems used to keep the optical elements clear of snow and ice, allowing continuous unattended operation in the Arctic conditions. In Section \ref{sec:obs} we describe the data collected during our test campaign, along with the data-reduction pipeline we developed for the systems. Section \ref{sec:perf} details the image quality and photometric performance of the cameras, and section \ref{sec:vars} describes some initial astrophysical results from the dataset. Section \ref{sec:disc} concludes with a discussion of concepts for much larger arctic surveys based on these camera prototypes.

\section{Transit Detection Efficiency at Ellesmere Island}
\label{sec:site}
Located on a 610-m ridge at the tip of the of the Fosheim Peninsula, the PEARL facility -- often referred to as the ``Ridge Lab'' -- is accessed by a 15 km long road from Eureka, a research base operated by the meteorological division of Environment Canada, the weather service of the Canadian government. The Eureka base is open to air traffic year-round, allowing astronomers to visit the site for observing runs throughout the winter.

Each year, the sun is continuously below the horizon at the Ridge Lab site for 3096 hours (129 days), and is continuously lower than $-12^{\circ}$ 1224 hours (51 days). Fall and spring ``shoulder seasons,'' when the sun is up for short periods, contribute a further 1,441 hours of dark time, for a total of 2,665 hours during which the sun is below $-12^{\circ}$ each year. The continuous darkness during the winter months allows continuous monitoring of a field, eliminates day aliases in the measurement of variable objects, and improves the detection probability of a repeating event such as a planetary transit \citep{Pont2005, Daban2010,Crouzet2010}.

\begin{figure}
  \centering
  \resizebox{1.025\columnwidth}{!}
   {
	\includegraphics{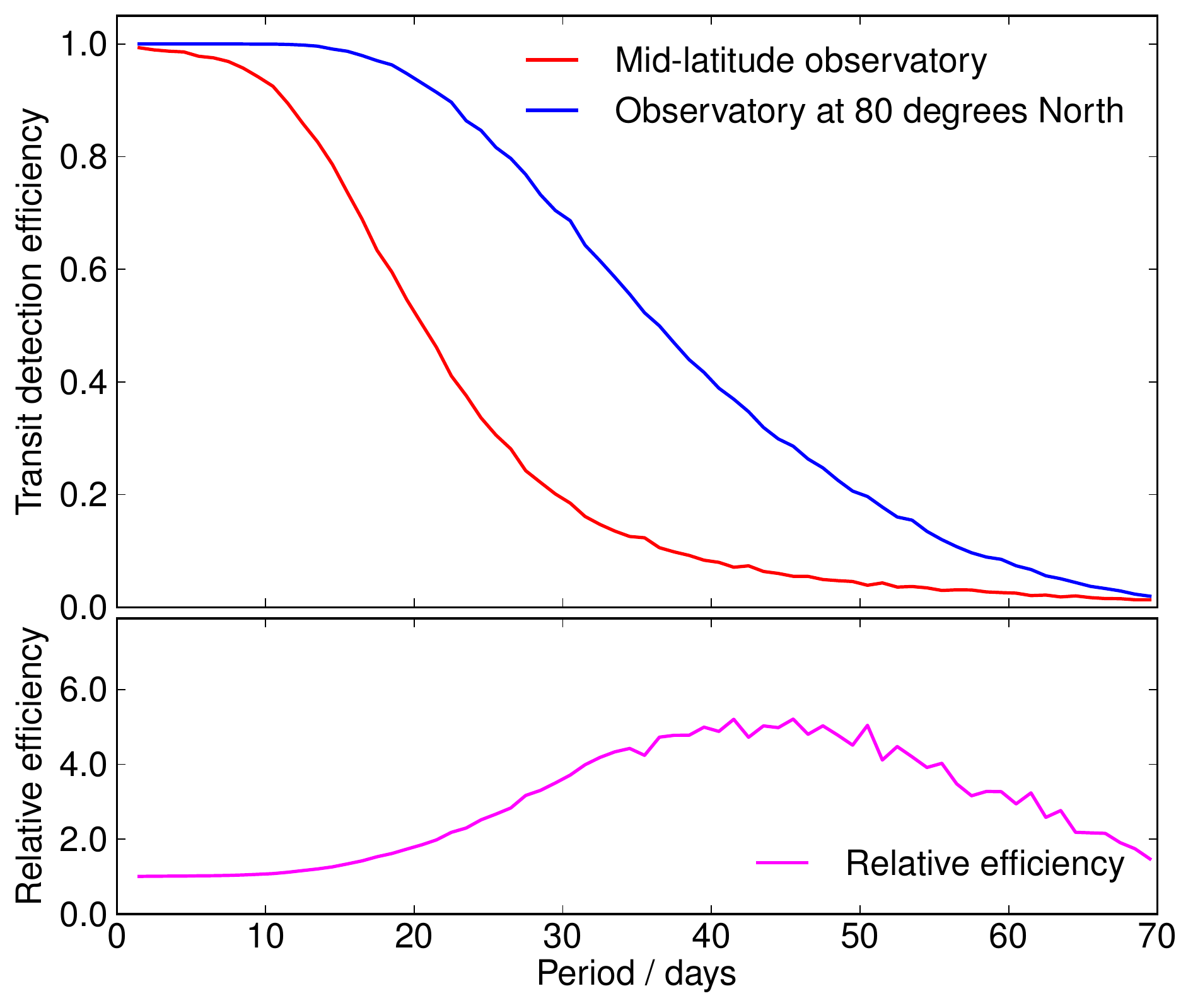}
   }
   \caption{A comparison of the probability of detecting a transiting planet in the field of a year-long survey, at a mid-latitude observatory and at an observatory at 80 degrees North. Because our exoplanet search is targeted at bright stars, we do not include the effects of moon brightness. The metric displayed here does not include the improved number of planet detections available if multiple fields around the sky are chosen for observation from the mid-latitude observatory; however, this cannot improve the numbers of long-period planets detected from mid-latitude sites by more than a factor of 2-3 at best. To remove the effects of aliasing, each point in the curves is the average efficiency over a period range of one day.}

   \label{fig:detect_effec}
\end{figure}

To quantify this improvement, we performed Monte-Carlo simulations of the detection efficiency of a transiting planet survey as a function of orbital period for both a mid-latitude site and the Ridge Lab site (Figure \ref{fig:detect_effec}). In the simulation, we assume that the telescopes only observe when the Sun is below an altitude of $-12^{\circ}$ and when the target is above an airmass of 2.0. We remove nights in which the available observation length is less than three hours, as is difficult to establish a useful photometric baseline for transit detection in that time. We include weather-time-loss in each simulation, with 24-hour periods of bad weather starting at randomly-chosen times (with overlapping bad-weather periods being allowed). We simulated correlated weather outages by assigning a 25\% probability for a bad-weather period being immediately followed by another period of bad weather (and so on for subsequent days). The weather simulation parameters were chosen to approximately reproduce typical weather outage patterns at mid-latitude sites (with a total of 35\% of observing hours affected by weather), but provide a conservative estimation for the Arctic site, which often sees periods of hundreds of hours of clear skies in the winter\citep{Steinbring2012}. For the secure detection of an exoplanet candidate we require the detection of three transits. 

As shown in figure \ref{fig:detect_effec}, Polar sites offer much improved detection efficiency for longer period planets, although we note that a mid-latitude site can improve the number of detected shorter-period planets by observing several fields throughout the year. Circumpolar surveys relying upon precision photometry or astrometry can also benefit from the near-zenith location of the celestial poles. At the Ridge Lab site an object at 80 degrees declination can be continuously tracked throughout the winter months at an airmass range of 1.00 - 1.06, a small range of motion that can reduce atmosphere-induced systematic errors for many astronomical programs. 

\section{Instrument Design}
\label{sec:design}
The AWCam prototype systems were required to operate autonomously in temperatures as low as -45$^{\circ}$C, to avoid or remove ice and snow deposition on their optical surfaces, and to survive storm events including blowing snow with wind speeds as high as 40m/s. For these reasons, we designed the systems to be both simple and extremely robust.

\begin{figure}
  \centering
  \resizebox{1.025\columnwidth}{!}
   {
	\includegraphics{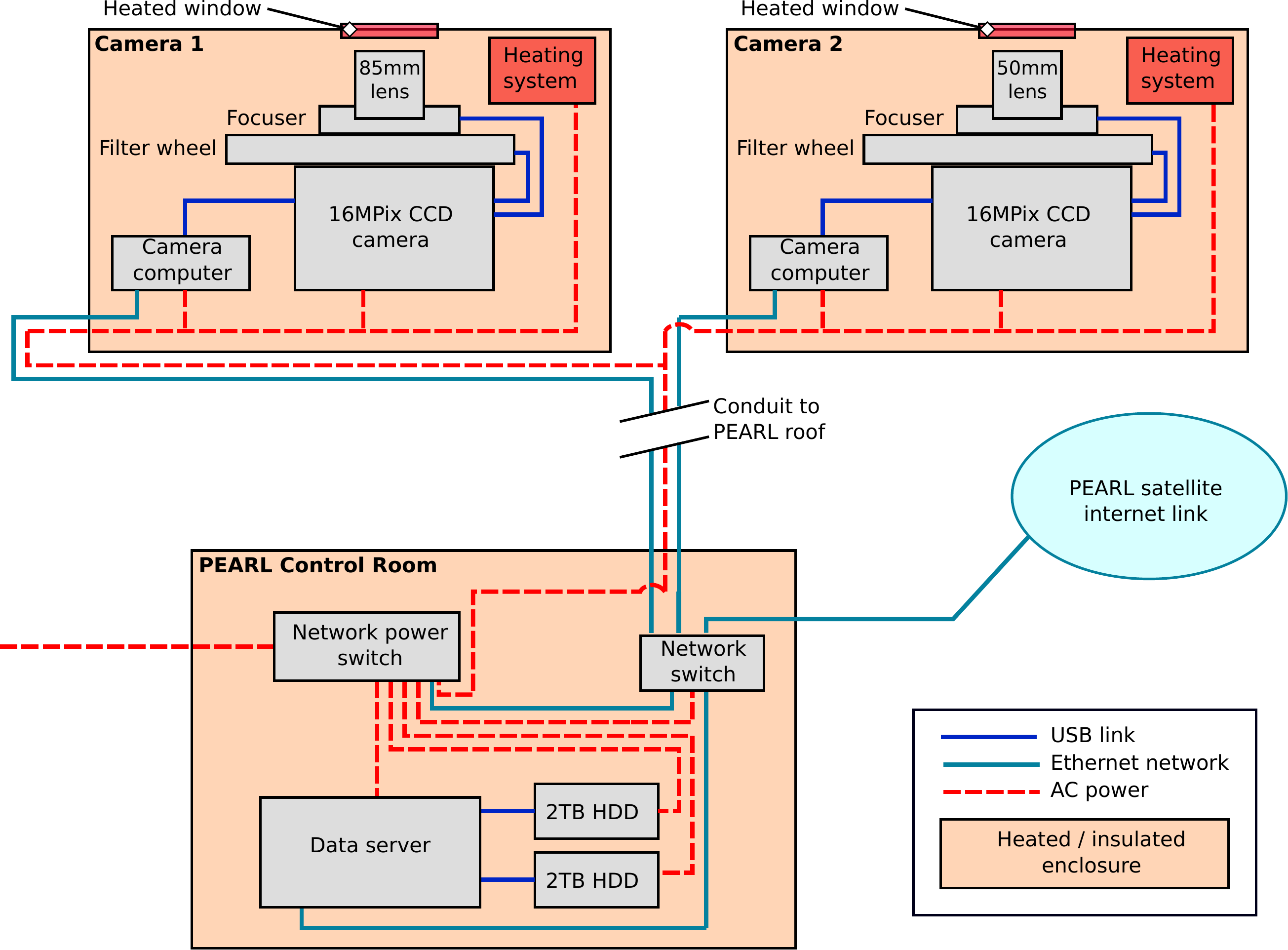}
   }
   \caption{A block diagram of the camera and data recording hardware. The cameras were placed in convenient locations on the Ridge Lab roof, with an approximately 100' cable run to the networking and data storage hardware in the PEARL control room.}

   \label{fig:block_diagram}
\end{figure}

The cameras are based on a fast Digital Single Lens Reflex (DSLR) camera lens mounted to a 16MPix CCD camera. The wide-field-imaging system is aligned on the North Celestial Pole, and we take short 10s exposures to avoid the need for sky-rotation tracking. Continuous photometry of a circular area of several hundred square degrees around the Pole can be obtained. In the 2012 campaign we tested two systems, identical except in their lens and filter choices. The hardware and survey characteristics are summarized in Table \ref{tab:survey_specs}, and an overview of the hardware is shown in Figure \ref{fig:block_diagram}.

The two camera systems were mounted on the observing platform of the Ridge Lab building, with Ethernet and AC-power connections to a warm control room within the building. The control room also contained a network power switch for remote rebooting of the systems, a network switch, and a server for data storage.

\subsection{Camera assembly}

Each camera system is based on an f/1.2 Canon EF-mount DSLR lens, with focal lengths chosen to provide fields of view of hundreds of square degrees, and sufficiently small pixels (in the tens-of-arcseconds range) to allow precision photometry of $\rm{m_V<10}$ stars without significant crowding effects. The lenses image onto Kodak KAF-16803 CCDs packaged in a Finger Lake Instrumentation (FLI) PL-16803 camera. A FLI CFW-4-5 filter wheel provides five filter positions for 50mm square filters. The Canon lens' built-in focus hardware cannot be easily accessed without connection to a Canon camera, and manual focusing of these very fast lenses did not provide sufficient image quality. To allow precision focusing under remote computer control we added an FLI Atlas focuser to the optical system. The camera systems are controlled over a USB link by a compact Linux-based server located inside their thermal enclosures. The completed imaging system is compact, measuring a total of 230$\times$230$\times$175mm and weighing 6.5kg.

\begin{table}
\caption{\label{tab:survey_specs}The specifications of the AWCam systems}

\begin{tabular}{ll}
\multicolumn{2}{l}{\bf Survey characteristics}\\
\hline
Pointing & North Celestial Pole\\
Survey dates & 14 February 2012 -- 21 February 2012\\
Survey length (total) & 152 hours\\
Survey length (dark and clear) & 98 hours \\
Data collected & 44,583 images (1.36 TB)\\
\vspace{0.1cm}\\

\multicolumn{2}{l}{\bf CCD Hardware}\\
\hline
CCD   & 4096$^2$ front-illuminated (KAF-16803)\\
Peak CCD Quantum Efficiency & 59\% \\
Pixel size & 9$\rm{\mu}$m \\
Readout time & 4s \\
\vspace{0.1cm}\\

\multicolumn{2}{l}{\bf 85mm camera}\\
\hline
Camera lens & Canon EF 85mm f/1.2L II USM\\
Field dimensions & 25.4 $\times$ 25.4 degrees\\
Continuous-coverage field  & 504 square degrees \\ 
Pixel scale        & 22.3"/pixel \\
Image quality        & 2-5 pixel FWHM over entire field \\
Filters              & Clear, g, r, i, z\\
\vspace{0.1cm}\\

\multicolumn{2}{l}{\bf 50mm camera}\\
\hline
Camera lens & Canon EC 50mm f/1.2L USM\\
Field dimensions & 40.8 $\times$ 40.8 degrees\\
Continuous-coverage field  & 1295 square degrees \\ 
Pixel scale        & 35.9"/pixel \\
Image quality        & 2-5 pixel FWHM over entire field\\
Filters              & Clear, g, r, i, z\\
\vspace{0.1cm}\\

\hline \vspace{0.25cm} \\
\end{tabular}
\end{table}

  \subsection{Enclosure, window and thermal system}
To ensure survivability of the electronics and other components, and to provide sufficient heating to remove snow and ice, the camera thermal enclosures were required to maintain an internal temperature of 5C in temperatures lower than $-40^\circ$C and in wind speeds up to 10m/s. We designed heated enclosures based on the Pelican 1610 case, an equipment shipping case which consists of a waterproof polypropylene copolymer enclosure. We drilled a 3.5-inch diameter hole on the upper surface for the camera aperture; a 3mm-thick AR-coated glass window protected the camera lenses. Two small holes equipped with bulkhead connectors provide inputs for AC power and Ethernet connections. The entire case internal surface was insulated by one-inch-think neoprene/vinyl/buna-N foam rubber.

A high level of interior temperature stability was required to provide sufficient focus stability for the small depth of focus of our F/1.2 lenses. We based the case thermal control system on the heating and air circulation system of a Dotworkz Systems D2-RF-MVP CCTV enclosure which was used in an earlier camera prototype. We modified the control electronics to allow control by an Omega CN743 precision temperature controller equipped with an Omega HTTC36-K-14G-6 hollow tube thermocouple probe. The heating system provides a total of 60W when fully activated, adding to the 70W produced by the cameras, computers and other electronics in the enclosures.

Pre-campaign testing of the systems was performed in a freezer at -27$^{\circ}$C, with wind simulation provided by small fans. We found the enclosure's thermal control systems could maintain temperature inside the cases to $\pm0.5^{\circ}$C in conditions approximating those found at the Ridge Lab site. On-site, the systems were able to maintain the nominal 5$^\circ$C internal temperature in all conditions encountered ($-30$$\pm$$5^{\circ}$C), except when the wind speeds exceeded 10m/s. 

\begin{figure}
  \centering
  \resizebox{1.0\columnwidth}{!}
   {
	\includegraphics{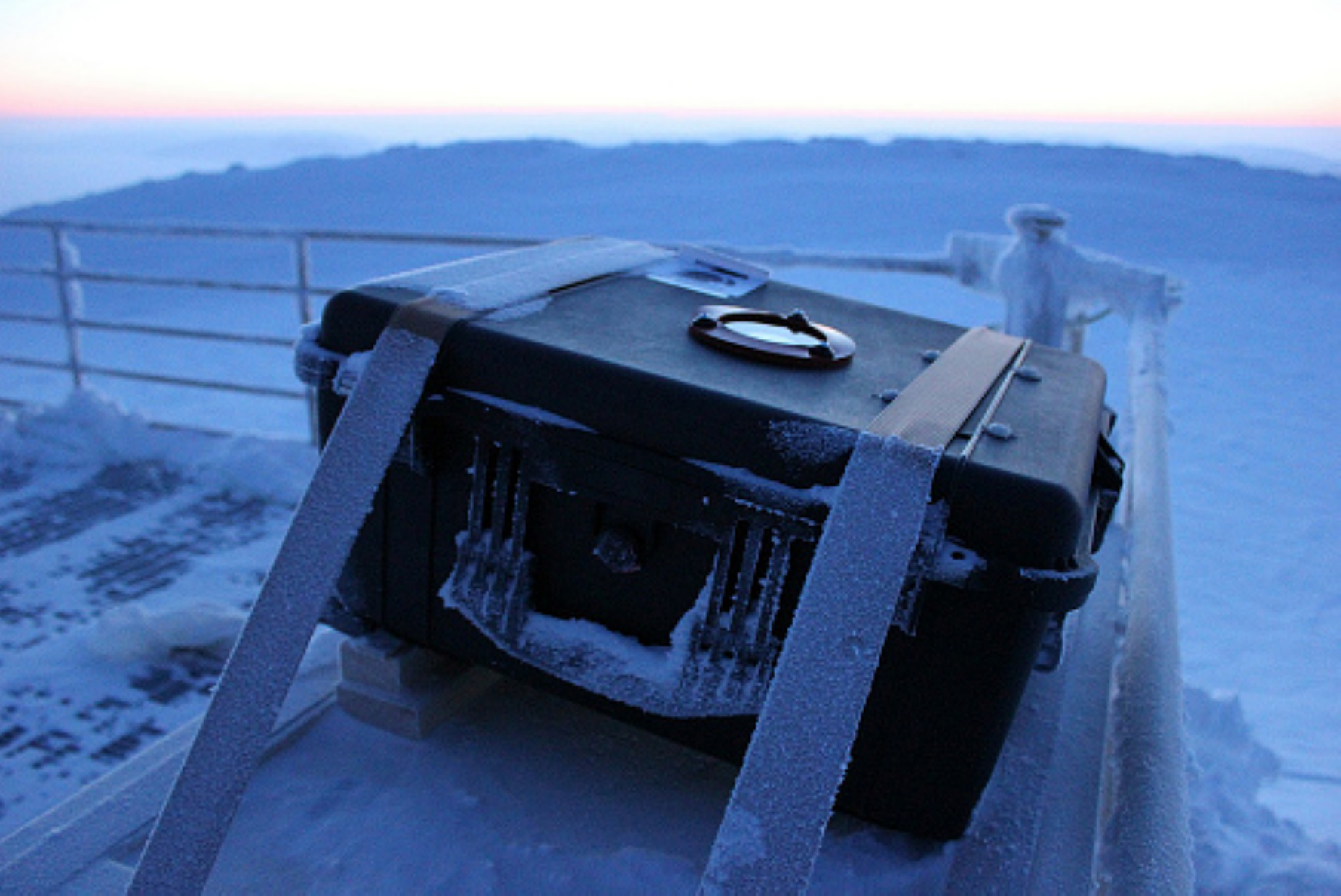}
   }
   \caption{One of the camera enclosures mounted at the Ridge Lab and aligned to the North Celestial Pole. Note the ice around the sides of the case and its surroundings, while the optical window is kept clear of accumulation by the heating system.}

   \label{fig:frost}
\end{figure}

Precision photometry requires keeping the optical window of the systems free of snow and ice during arctic conditions. Our testing in the Arctic, including days with high winds, blowing snow and high levels of ``diamond-dust'' (fine ice particles) demonstrated that the window was kept clear by the heating system in all conditions encountered (Figure \ref{fig:frost}). The gentle heating allowed the snow and diamond dust to sublimate off the camera enclosures rather than melting, reducing ice buildup around the instruments.

  \subsection{Camera Mounting and Alignment}

The cameras were aligned on the North Celestial Pole by rotating the enclosures to face North and tilting them to the required angle. An initial azimuthal alignment was measured using a North-South line pre-determined via a handheld GPS receiver, and the system tilt was fixed using a digital inclinometer. After initial pointing, fine alignment was achieved by monitoring the position of the North Celestial Pole in the acquired images. We used ratchet straps and shims to mount the enclosures onto the grillwork on the Ridge Lab roof, a design which proved stable for all the encountered conditions.

  \subsection{Camera Control Software \& Data Storage}

The camera systems were controlled by custom software running on Ubuntu-Linux-based computers inside each enclosure. The software handled initial hardware tests, autofocusing, routine camera operations,  and automatic transfer of acquired data to the data server.

The autofocus routine was based on code used to focus the Palomar Transient Factory camera \citep{Law09}. First, the camera steps through a range of focus positions. Star-like objects in the field are extracted and have their FWHMs measured using SExtractor \citep{Bertin1996}. The resulting list of objects is then filtered to exclude extended objects, cosmic rays and other non-point-sources, a median FWHM is derived for each focus position, and a parabola is fit to the resulting focus position / FWHM curve. The DSLR lenses produce a changing PSF across the field (see section \ref{sec:fwhms}), including field curvature which can interfere with simple autofocus routines because different parts of the field can be in focus at different positions. We found that restricting the focus routine to use only the central 80\% of the image and rejecting stars with high ($>$1.5) ellipticity gave acceptable focusing for targeted zone of continuous coverage in the center of the field.

 The two camera systems each generated approximately 8GB of image data per hour. We transferred the data immediately after acquisition onto two 2TB hard disks inside the Ridge Lab. The satellite Internet connection at the site was sufficient only for test and monitoring images to be transferred during observations; the full survey dataset was stored on USB external hard disk drives for later analysis.

\section{Observations and Data Reduction}
\label{sec:obs}

Table \ref{tab:obs} details the data taken during our February 2012 campaign.  Because the campaign was undertaken towards the onset of arctic sunrise, twilight conditions were encountered for a fraction of the day, allowing a total of 16 hours of dark-sky observing in each 24-hour period. The cameras operated continually from 14 February 2012 to 21 February 2012, a total of 152 hours of which 98 hours was dark. Thin-to-moderate clouds were experienced for approximately half the run, resulting in 0.5-2 magnitudes of extinction and increased scattering during those periods.

To test different survey methods, we operated the two cameras systems in slightly different ways. The 85mm camera swapped sequentially between five filters between each exposure, providing near-simultaneous multi-colour photometry. The 50mm camera instead operated in a single fixed r-band filter. The r and i bands of the 85mm camera had poor image quality due wedge angles in the filter glass; we continued observations in those filters for sky-brightness measurements only.

The cameras operated with an open-shutter efficiency of 71\% for the 50mm camera and 65\% for the 85mm camera, which had slightly increased overheads because of its filter changes. In total, the cameras collected 44,583 images, \mbox{1.36 TB} of data. 

\begin{table}
\caption{February 2012 observations\label{tab:obs}}

\begin{tabular}{lcccc}
\bf{Camera} & \bf{Filter} & \bf{Total images} & \bf{Open shutter} & \bf{Cadence}\\
\hline
85mm lens & Clear    & 4575               & 12.7 hours & 77 secs\\
                    & g           & 4580               & 12.7 hours & 77 secs\\
                    & r           & 4621               & 12.8 hours & 77 secs\\
                    & i           & 4580               & 12.7 hours & 77 secs\\
                    & z           & 4578               & 12.7 hours & 77 secs\\
\\
50mm lens & r           & 21649               & 60.1 hours & 14 secs\\
\vspace{0.2cm}
\end{tabular}
\end{table}

\begin{figure*}
  \resizebox{1.0\textwidth}{!}
   {
	\includegraphics{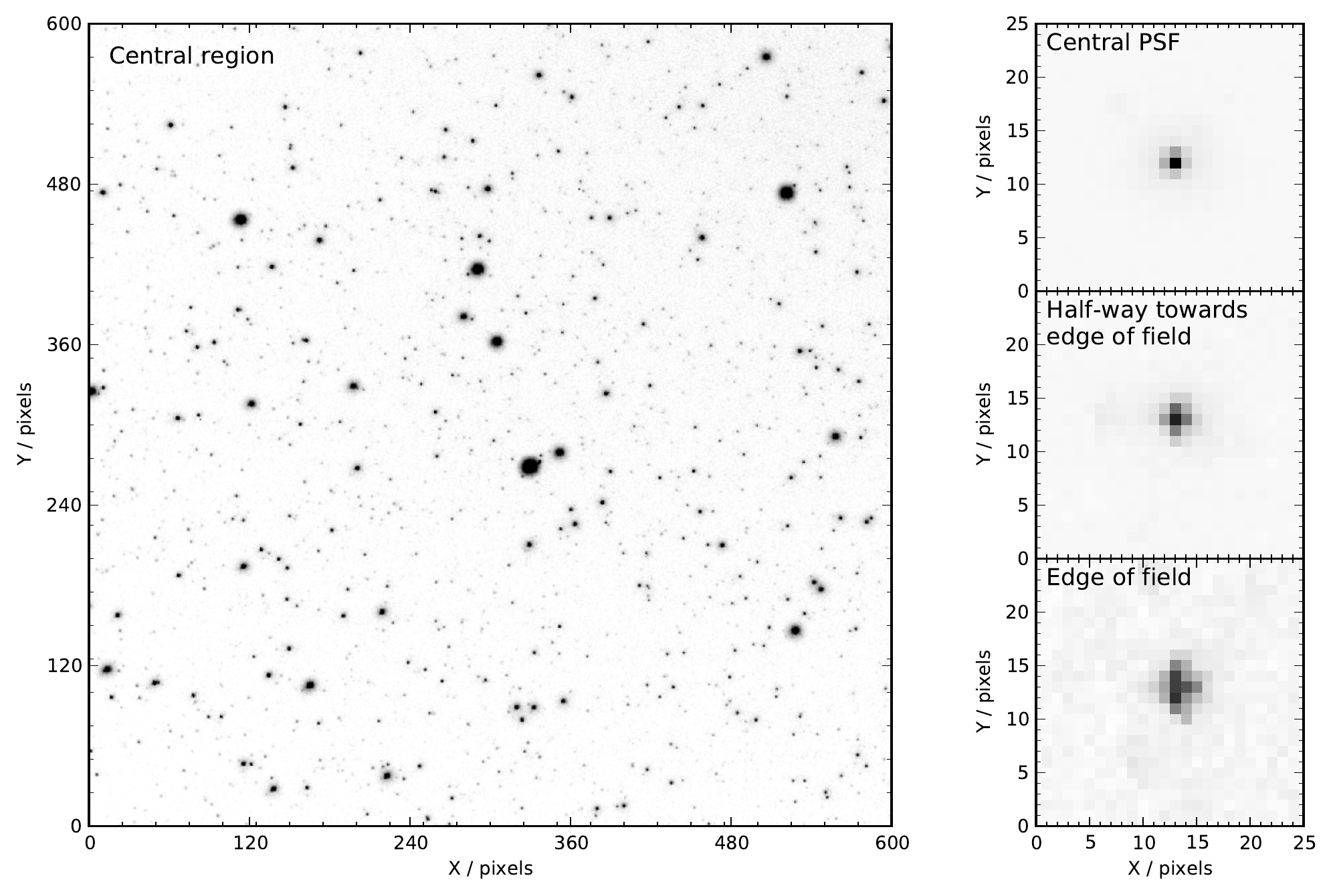}
   }

   \caption{Image cutouts from the 85mm-lens camera system, a 10s exposure in the g filter. \textit{Left:} a central region from the image; \textit{Right, top-to-bottom:} three small cutouts showing representative PSFs at the center, halfway to the edge, and the edge of the field respectively.}
   \label{fig:psfs}
\end{figure*}

  \subsection{Flat fielding with very-wide-field cameras} 
  
Flat fielding with a fixed, autonomous camera system is potentially challenging. We tested four flat fielding methods to inform subsequent instrument and survey design -- conventional twilight flats, superflats generated from our full dataset, imaging a flat white screen held above the cameras, and using thick clouds as flatfields.

\subsubsection{Conventional twilight flats}
Flat fields taken during twilight are susceptible to large-scale gradients caused by multiple scattering. These effects can generally be neglected for narrow-angle astronomical imaging, as long as extremely high precision is not required. In contrast, our very-wide-field twilight flats showed unacceptably large gradients. This, coupled with the several months during which twilight is not available at circumpolar observing sites, makes twilight flats unsuitable for these camera systems. 

\subsubsection{Superflats}
The major contributors to the sky brightness measured by superflats generated from data taken in dark time are airglow, zodiacal light, and starlight\footnote{The Ridge Lab is located near the North Magnetic Pole, ensuring that most auroral emission is over the Southern horizon (Figure \ref{fig:map}).} Zodiacal light and airglow vary across the sky; at a zenith angle of 12.5 degrees (the edge of our narrowest field) the airglow is approximately 2.5\% brighter than at the zenith, while the zodiacal light varies by approximately 10\% across the range of ecliptic latitude covered by our cameras \citep{Benn1998}. Because of our cameras' large pixels, the starlight contribution is larger than found in narrow-field astronomical images, and the repetitive nature of the sky rotation complicates the generation of precision superflats. For these reasons, we found that it was extremely difficult to generate precision superflats from our dataset.

\subsubsection{Screen flats}
``Dome'' flats generated from screens in front of the camera are not susceptible to the problems encountered by twilight and super flats, although care is required to ensure a uniform illumination pattern. Achieving this is simple for our fast, small lenses. We found that a white card held in front of our cameras during late twilight was sufficient to generate high quality flats (at a level sufficient for millimag photometry) for both our cameras and all filters.

\subsubsection{Cloud flats}
Without a dome on which to place the screen, screen flats cannot be easily taken by a standalone remote camera system. An automated system to place screens over the cameras would be complex and potentially unreliable. For these reasons, we also attempted to test the use of thick clouds as a substitute flat screen. The thickest clouds during which we obtained data, with an extinction of $\sim$3.5 magnitudes, still had enough stars to complicate the generation of flat fields. Comparing the "cloud-flats" to the screen generated flat, we also found variations in the sky brightness at the 1\% level on scales of 5-10$^{\circ}$, likely induced by structures within the cloud layer. This could possibly be removed by averaging of a long-term cloud-covered dataset, but the cloudy data we collected was not of sufficient length to test this.

\subsubsection{Reliable Wide Field Camera Flats}
Twilight and cloud-based flats appear to only be suitable for rough flat-fielding with these wide-field camera systems. Although the sky-gradient effects of scattering, airglow and zodiacal light could removed at some level using their predicted across-field variations, it is difficult to generate clean flats with the stars in the field executing repetitive motions that puts them in the same pixels at the same time each day. Cloud-generated flats showed percent-level spatial changes that will be difficult to accurately remove without a long period of cloudy weather, and there is a possibility that the brightness distribution of low-level clouds could be systematically biased by local topographic features. 

For these reasons, we elected to use the screen flats for subsequent analysis. Remote operation of these systems is challenging, although we note that the sealed optical system of our cameras and their very fast beams reduce the likelihood of changes in the flats. We did not detect any flat field changes over the course of our survey, at the one part in 1,000 level; it is likely that an occasional visit by a technician placing illuminated screens in front of the cameras would be sufficient to achieve precision flat-fielding.

\subsection{Data reduction pipeline}
\label{sec:pipeline}
    We implemented a Python-based data reduction pipeline to calibrate our images for photometry and astrometry. FITS files recorded by the cameras were first dark-subtracted and then flat-fielded using the methods described above. The sources in each field were extracted using SExtractor \citep{Bertin1996} ``mag-auto'' apertures with modelling and subtraction of a background that was allowed to vary across the field.

\subsubsection{High-efficiency astrometric calibration of extremely-wide-field images}
\label{sec:astrom}

The images produced by our lenses have a generally much higher degree of geometric distortion than those produced by standard astronomical telescopes. We found that automatic astrometric calibrations typically used for wide-field surveys such as astrometry.net \citep{Lang2010} and SCAMP \citep{Bertin2006} led to unacceptably poor astrometric solutions ($\sim$$100$ arcsecond median astrometric error compared to catalogs in some regions of the image), even with high-order polynomial corrections. We found that splitting the individual images into 10-30 rectangular subsections allowed SCAMP to find an acceptable astrometric solution within each subsection based on the UCAC-3 catalog \citep{Zacharias2010}. However, each subsection required tens of seconds to extract the sources and find an astrometric solution. This results in whole-image processing times of several minutes, unacceptable for images taken at least 4 times each minute in two cameras.

To improve the processing times we developed a model to directly relate camera pixel positions and observation times to sky coordinates. Because the cameras' optical axes were slightly mis-aligned with respect to the Celestial Pole, we modelled lens distortion terms in polar coordinates around one ``optical'' axis, and then the conversion to RA and Declination around a separate ``celestial'' axis. We modelled the lens distortion in the following way, based on the models first described in \citet{Conrady1919} and \citet{Brown1966}:

\begin{equation}
D = \sqrt{(x-x_c)^2 + (y-y_c)^2}
\end{equation}

\begin{eqnarray}
x_{lens\, corr} &=& (x-x_c)(1 + aD^2 + bD^4 + cD^6) \nonumber\\&& + 2d(x-x_c)(y-y_c) + e(D^2 + 2(x-x_c)^2))
\end{eqnarray}

\begin{eqnarray}
y_{lens\, corr} &=& (y-y_c)(1 + aD^2 + bD^4 + cD^6) \nonumber\\&&+ d(D^2 + 2(y-y_c)^2 + 2e(x-x_c)(y-y_c)) 
\end{eqnarray}

where $x,y$ are the measured stars positions in pixels, $x_c, y_c$ are the camera optical axis pixel coordinates, and the terms a-e are fit to the measured astrometry. To measure the model coefficients and optical axis coordinates we took 20 frames distributed throughout the nights, fully calibrated the frames using astrometry.net and SCAMP, and used a downhill simplex minimization to fit the coefficients to the calibrated astrometry. After correction of the pixel coordinates, the sky coordinates of the stars are calculated by converting the de-distorted pixel positions to sky coordinates with a tangent plane projection and then adding the sky rotation term corresponding to the observation epoch.

After initial modelling, we found that a fixed lens model did not adequately fit the data. An additional shift was required to model a slow change in the angle between the optical and celestial axes. The amplitude of the modelled shift is consistent with a 100$\mu$m height change in one edge of the camera support structure, and was maximal around midnight. We attribute this to the effects of changing temperature during the semi-twilight nights. 

Verifying the full astrometric solution against a second set of calibrated images, we found that just based on the time of observation and the pixel position of the stars, sky positions were calculated to $\approx$20 arcsec ($<$1 pixel) accuracy across the entire image, with improved performance in the image centers. The full astrometric processing for the $\sim$10,000 stars in each image takes a fraction of a second.

\begin{figure}
  \centering
  \resizebox{1.0\columnwidth}{!}
   {
	\includegraphics{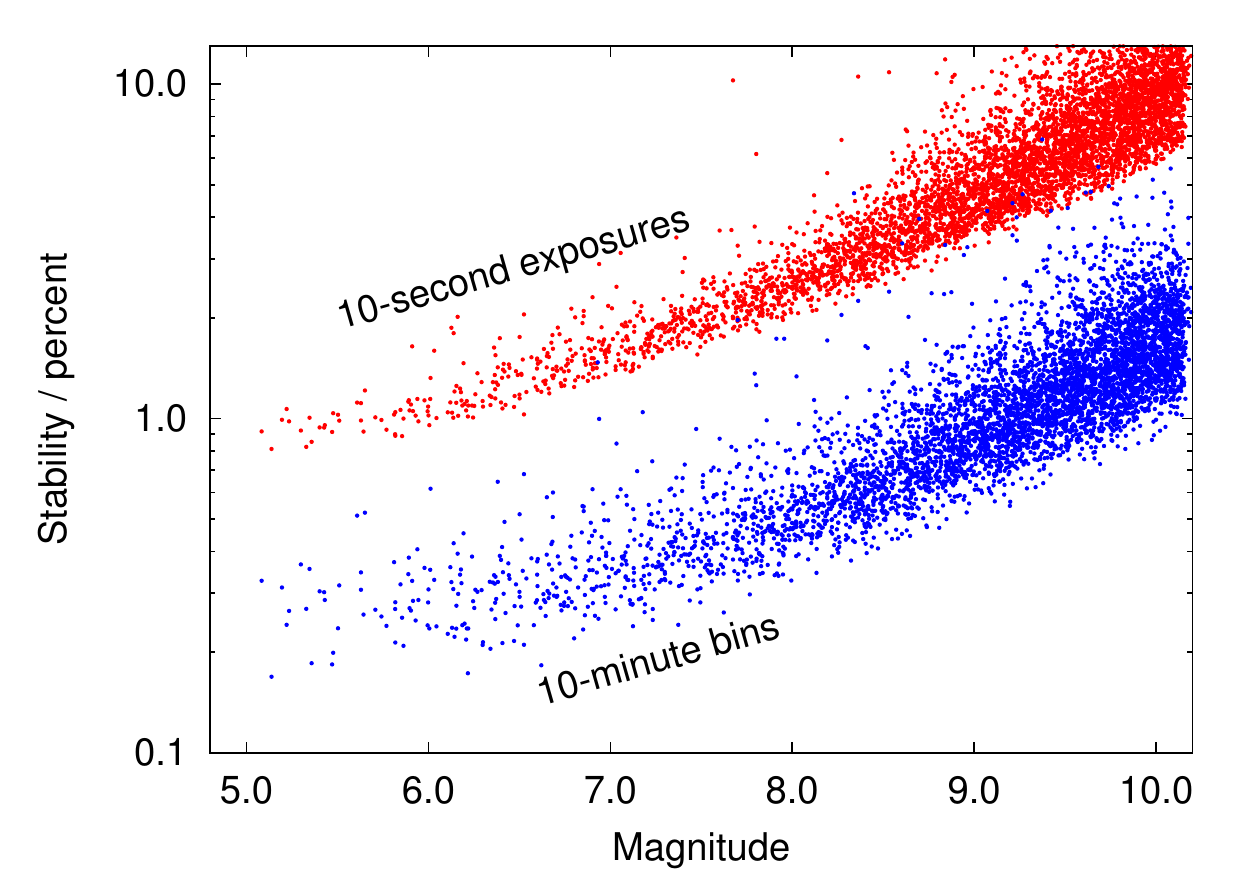}
   }
   \caption{The photometric precision achieved for the 5,283 bright stars in the 50mm camera field of view. The red points show the RMS variations in the measured photometry across the 2801 individual photometric data points taken for each star over 12 hours of operations. The blue points show the stability when binned down to 10-minute (60 datapoint) chunks.
   }

   \label{fig:full_phot}
\end{figure}

\subsubsection{Efficient millimagnitude photometry with extremely-wide-field images}

To obtain photometric zeropoints we modified the differential aperture photometry pipeline originally developed for the Palomar Transient Factory \citep{Law09} M-dwarf planetary transit search \citep{Law2012}. We associate an initial ``rough-guess'' photometric zeropoint with each image, and using them generate light curves for each source. We then vary the zeropoints, using a simulated-annealing algorithm to minimize the averaged noise-weighted RMS variability in all the light curves. The algorithm converges on the set of zeropoints which provide the most stable photometry for the largest number of sources in the field, and is thus robust to bias from astrophysical or instrumental variability in individual stars.

The slight mis-alignment between the cameras' optical axes and the Celestial Pole (Section \ref{sec:astrom}), and the small focal plane tilt (Section \ref{sec:fwhms}), leads to a slow change in size and shape of the stars' PSFs as the sky rotates. We found that this was enough to induce few-percent-level changes in the post-zeropoint-calibration stellar fluxes, for stars at the edges of the field where the PSF changes were largest. The resulting systematic flux error affects each star in a particular region of the image in the same manner, and so can be removed with standard methods designed to remove light curve systematics. We use 15 iterations of SysRem \citep{Tamuz2005} to model and remove this systematic error for the ensemble of light curves.

After photometric calibration and systematics removal, we consistently achieved 1-percent-level photometric performance in individual images over 12-hour timescales, and few-millimagnitude-level photometry when binning datapoints from 60 individual exposures together (Figure \ref{fig:full_phot}). 

The expected single-exposure scintillation noise for at-zenith observations is approximately 6 and 8 millimagnitudes for the 85mm and 50mm lenses respectively \citep{Dravins1998, Young1967}.  The scintillation noise estimates depend on the particular observing conditions at the site (to within $\sim$50\%; \citealt{Young1967}), and we thus conclude that the cameras are achieving very close to scintillation-limited photometric precisions. We evaluate the photometric performance for planet-search purposes, including the level of correlated noise, in section \ref{sec:phot_perf}.

\begin{figure}[]
  \centering
	\subfigure{\resizebox{0.49\textwidth}{!}{{\includegraphics{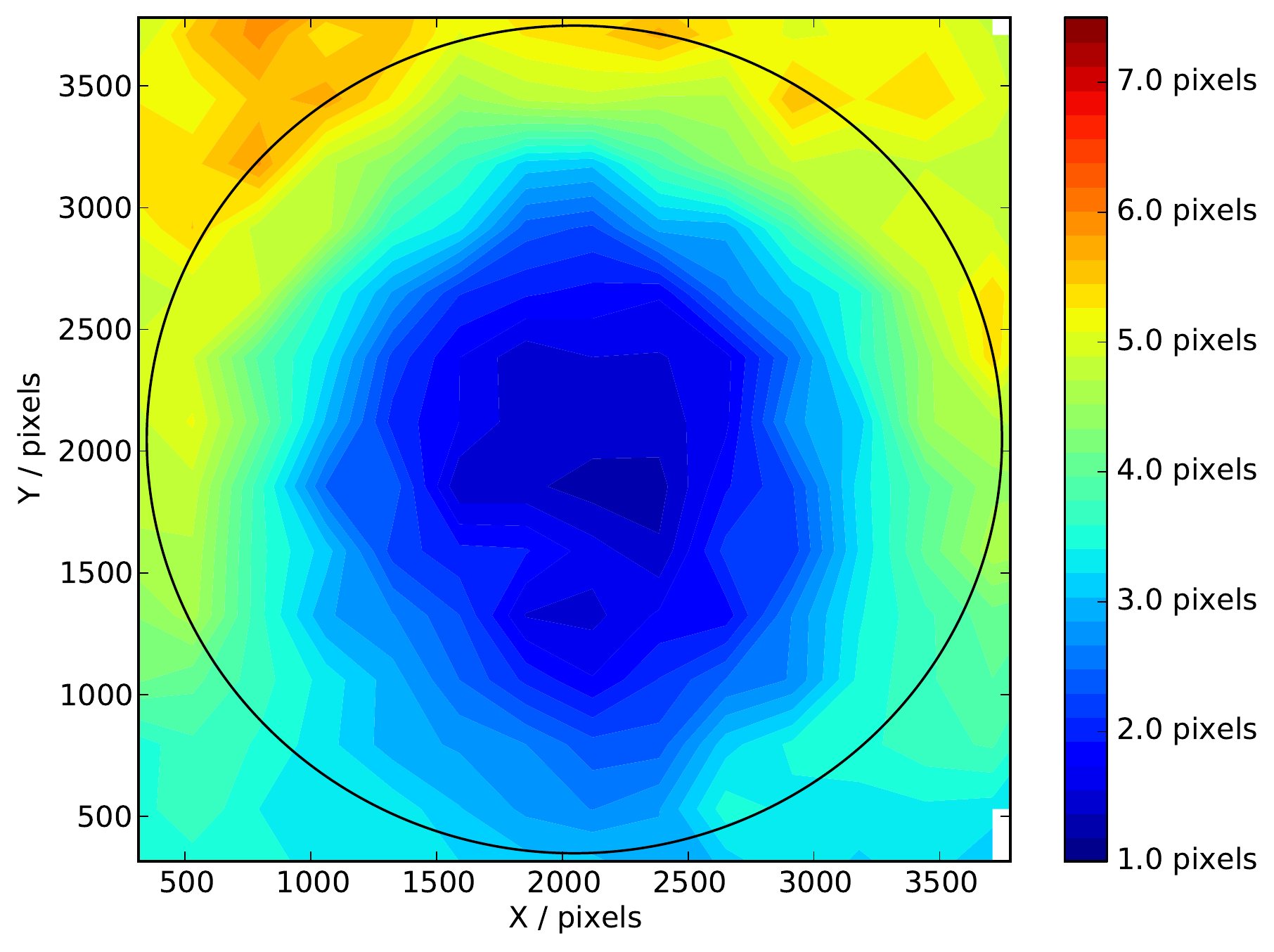}}}}
	\subfigure{\resizebox{0.49\textwidth}{!}{{\includegraphics{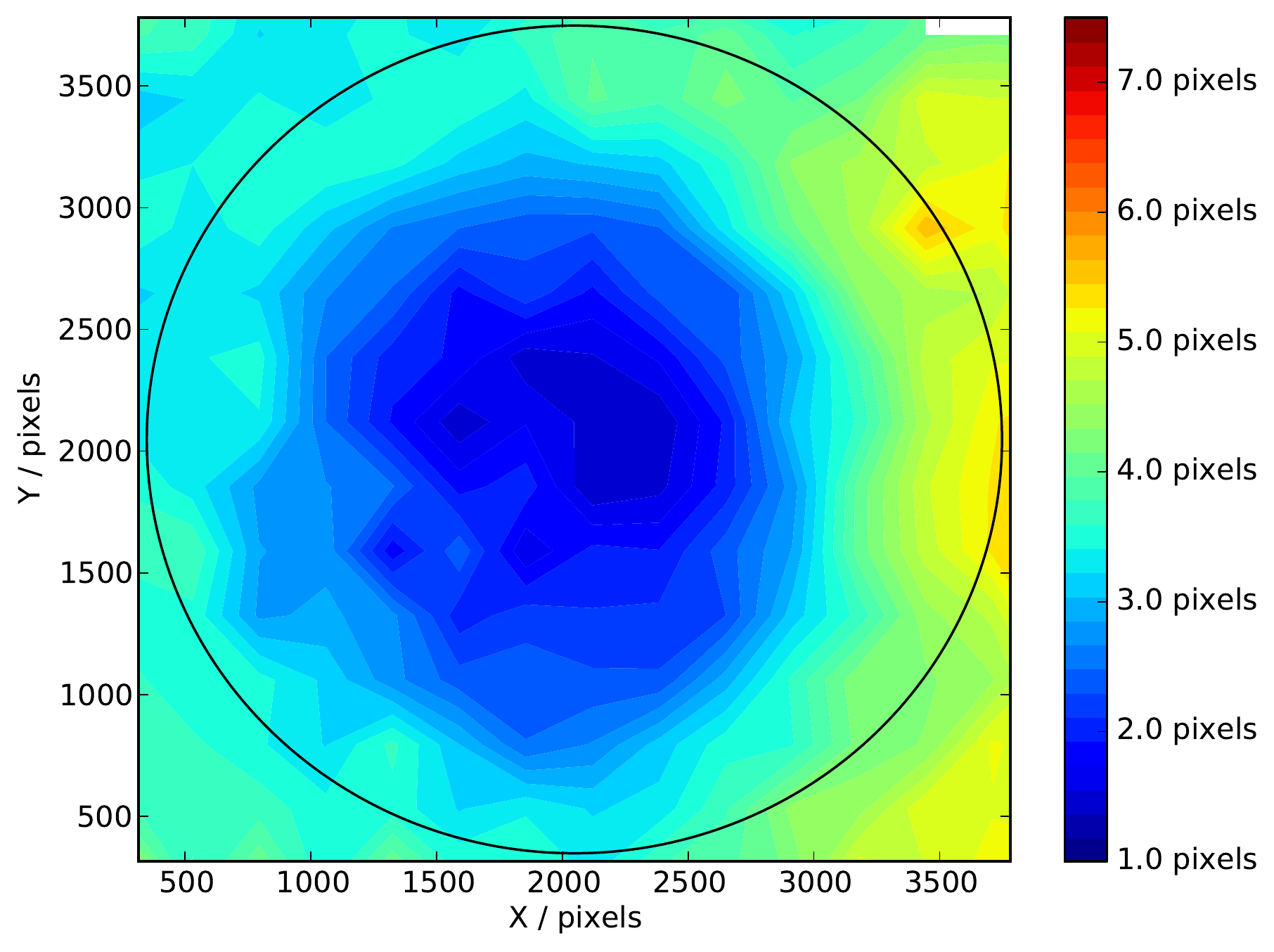}}}}
   \caption{Maps of the FWHM of stars recorded in the standard survey bands for the two cameras: the g filter using the 85mm F/1.2 lens (\textit{top}) and the r filter with the 50mm F/1.2 lens (\textit{bottom}). The circle marks the zone for which the camera PSFs were optimized, where stars have continuous photometric coverage as the sky rotates.}
   \label{fig:fwhm_maps}
\end{figure}

\section{Performance}
\label{sec:perf}

In the below subsections we detail the performance of the camera systems in the context of the requirements of a bright-star transiting exoplanet survey.

\subsection{Image Quality}
\label{sec:fwhms}
The lens point spread functions (PSFs; Figure \ref{fig:psfs}) are well matched to the camera's pixel size in the center of the field, with round shapes and full width at half maxima (FWHMs) of approximately two pixels. In both cameras, the PSFs increased in width towards the edges of the field (Figure \ref{fig:fwhm_maps}), although at a slow enough rate to provide useful images up to the edges of the zone in which stars are continuously covered. A small tilt in the focal plane ($\sim$5$\mu$m across the CCD) is visible in each camera's FWHM map, produced by a combination of the camera mounting hardware and the tolerance in the wedge angle of the filter glasses. The tilt did not significantly affect the image quality within the continuous coverage zone.

\subsection{Camera Sensitivity}

The cameras achieved their design sensitivity, with a typical 10s-exposure 5-sigma point-source limiting magnitude of $\rm{m_g}=12.6$ for the 85mm lens. We measured the limiting magnitudes using stars observed in a Sloan Digital Sky Survey SEGUE field \citep{Yanny2009} which was included in both our cameras' fields of view and allows us to approximately match filter responses for photometric calibration. At its highest point, the SEGUE field covers declinations from 82.3$^{\circ}$ to 85.1$^{\circ}$. Our cameras have somewhat lower image quality and increased vignetting at those altitudes than at the center of the fields. Correcting for these effects, the true limiting magnitude at the field centres is improved to $\rm{m_g}\approx13.6$.

\begin{figure}
  \resizebox{1.025\columnwidth}{!}
   {
	\includegraphics{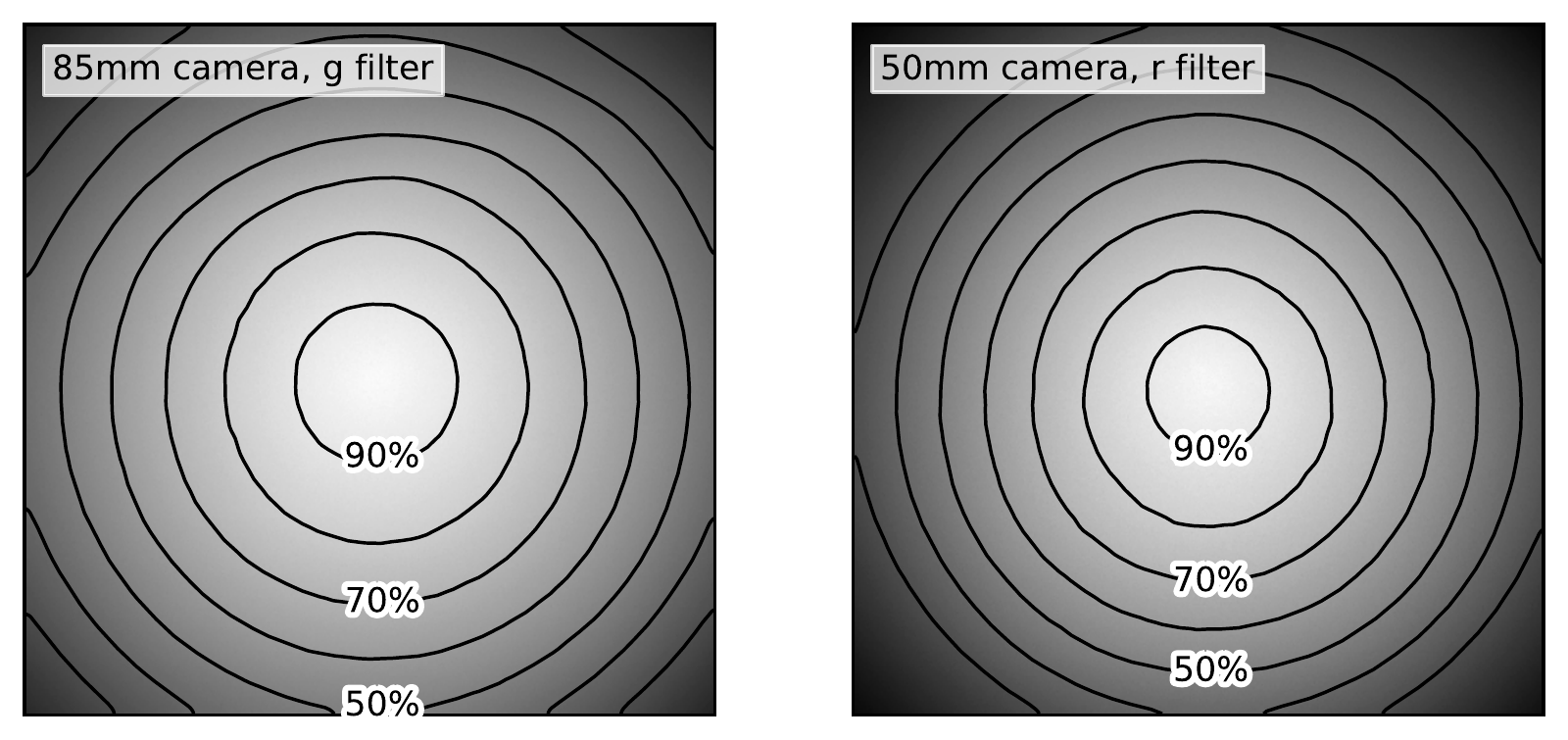}
   }

   \caption{The vignetting of the fields of the 85mm (left) and 50mm (right) lenses: the flatfield images from the cameras. A sensitivity of at least 50\% of the maximum value is maintained throughout the continuous-coverage zone.}
   \label{fig:sens}
\end{figure}

Figure \ref{fig:sens} shows the relative vignetting of our cameras as a function of field position. The 85mm lens reduces the light collection to approximately 50\% of its maximum value by the edge of the camera's zone of continuous photometric coverage, while the 50mm lens has a more pronounced vignetting giving $\approx$35\% sensitivity at the edges of the chip.

\begin{figure}
  \centering
  \resizebox{1.0\columnwidth}{!}{\includegraphics{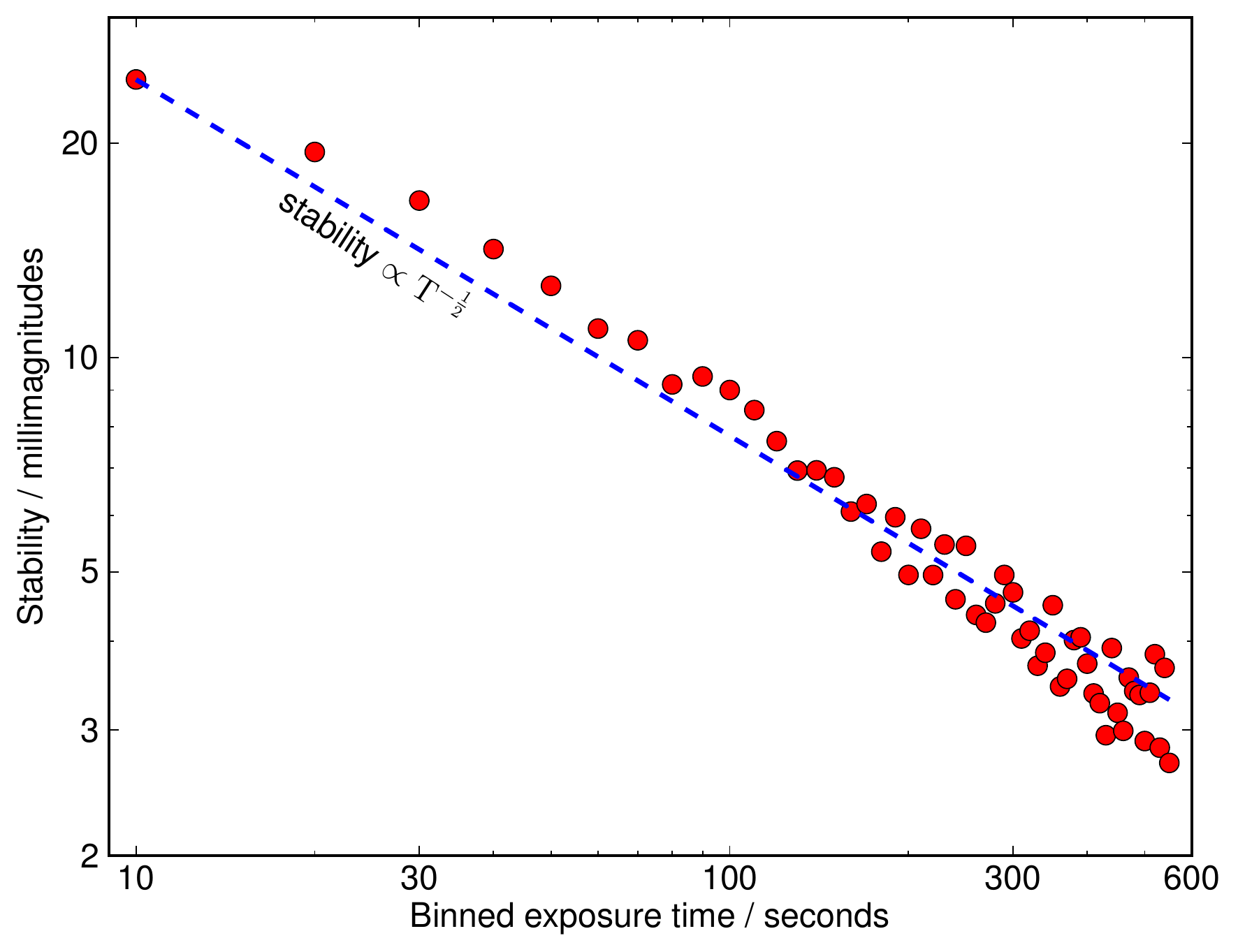}}
  \resizebox{1.0\columnwidth}{!}{\includegraphics{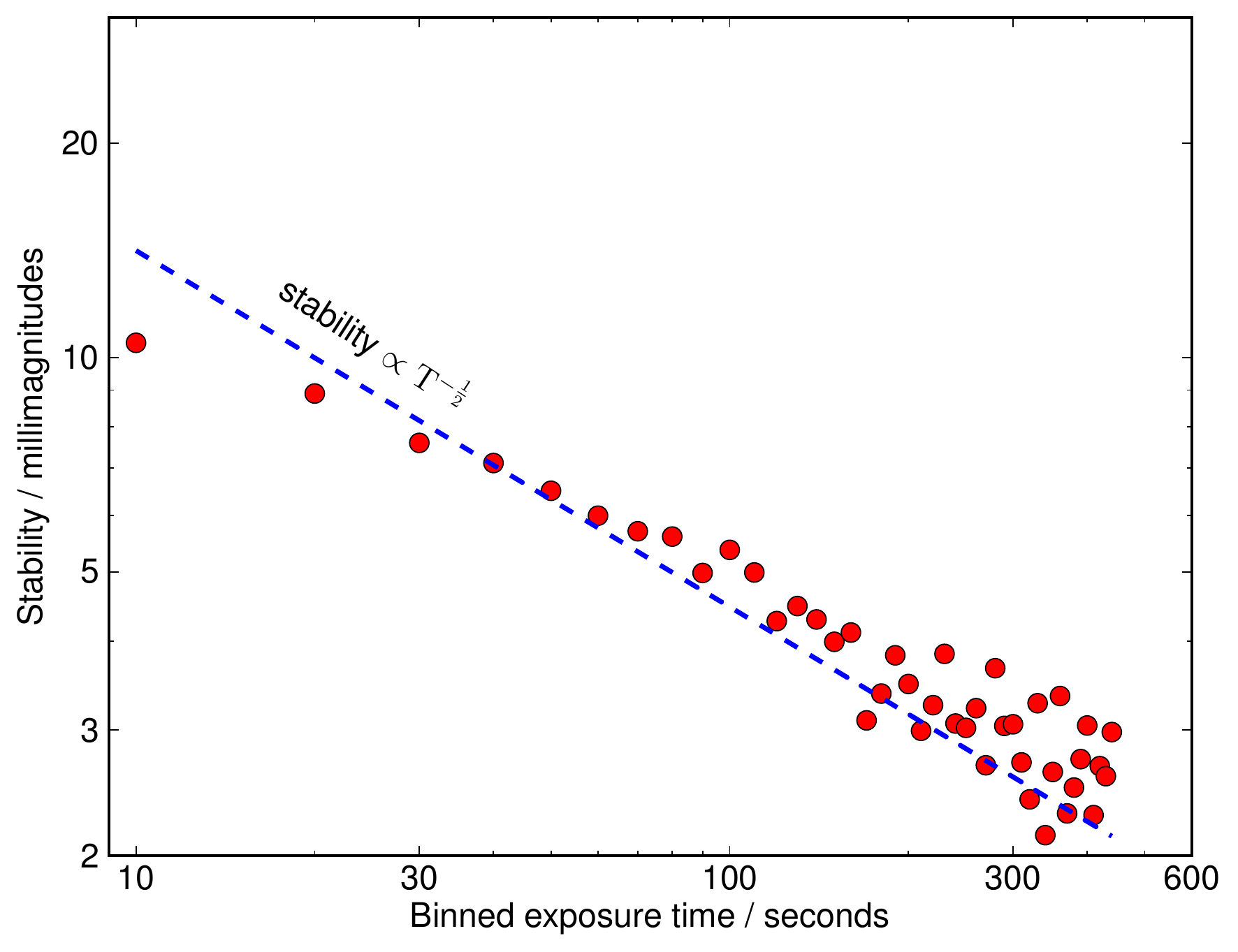}}
   \caption{The photometric stability (RMS compared to median magnitude) achieved by the 50mm camera (\textit{top}) in the r filter and the 85mm (\textit{bottom}) camera in the g filter on a typical $\rm{m_V=7.0}$ star, which is bright enough for its photometry to be limited by scintillation and systematics. We binned individual 10s-exposure points to achieve 2-3 millimag stability over effective exposure times of 5-10 minutes. The expected behaviour as exposure time increases, with the photometric precision improving with the square root of the total exposure time, is a close match to the measured performance.}

   \label{fig:phot_prec_bin}
\end{figure}

\subsection{Correlated noise performance}
\label{sec:phot_perf}
As shown in Figure \ref{fig:full_phot}, binning our single-exposure photometry allows few-millimagnitude precision for thousands of stars in the our fields of view. However, correlated (``red'') noise can greatly reduce the efficiency of planet detection even with apparently good photometric stability (e.g. \citealt{Smith2006}).

We evaluated the correlated noise performance of our camera systems by searching for unexpected behaviour when binning the photometric data points to different timescales; a system with a high degree of correlated noise would show a precision improvement with binning which is slower than the expected $\rm\sqrt{N_{data\, points}}$. To remove the possibility of obtaining spuriously good performance estimates by over-fitting during the systematic noise reduction steps, we used a subset of our full data reduction pipeline, without the final photometry and systemic noise removal steps. We repeated the below tests with our full pipeline and obtained identical results.

Using the SExtractor extraction of the sources in our astrometrically-calibrated images, we performed differential aperture photometry for four isolated stars near the center of our fields. We selected the target and three calibration stars to have magnitudes in our exoplanet survey range ($\rm{m_V=7-9}$), but with care to keep their peak fluxes below $\sim$$20,000$ ADUs to avoid non-linearity due to our CCD's anti-blooming system (which engages at signals of $\rsim$32,000 ADUs). Within each night, we fit a second-order polynomial to the light curves to remove long-timescale residual systematic variations (typically at the 5-millimagnitude level). The final photometric stability was measured from the RMS variability in the resulting light curve.

Similarly to the full pipeline, we achieved an RMS precision of 1.0\% in individual 10s exposures on the 85mm camera, and 2.4\% in r-band exposures with the 50mm camera. The ratio of photometric performance for the two lenses is roughly as expected for measurements limited purely by photon and scintillation noises, suggesting only low-level extra sources of photometric error.

Binning the photometry to 10-min exposures improved the photometric precision to 2-3 millimagnitudes (Figure \ref{fig:phot_prec_bin}), compared to an expected scintillation-limited precision of $\approx$2 millimagnitudes. In both cameras, the photometric precision improves the expected rate of  $\rm\sqrt{N_{data\, points}}$ until at least 100 frames are binned together. The length of our dataset from this campaign precludes a meaningful analysis of the effects of correlated noise on longer timescales, but the close adherence to the $\rm\sqrt{N_{data\, points}}$ relation is encouraging.

\section{Detected variable sources}
\label{sec:vars}
We performed a search for variable stars and other astrophysically varying sources, both visually searching for obvious periodic sources, and testing automatic variable-source detection in a blind search for stellar variability and transit signals.

\subsection{Bright periodic sources}
\begin{figure}
  \centering
  \resizebox{1.0\columnwidth}{!}
   {
	\includegraphics{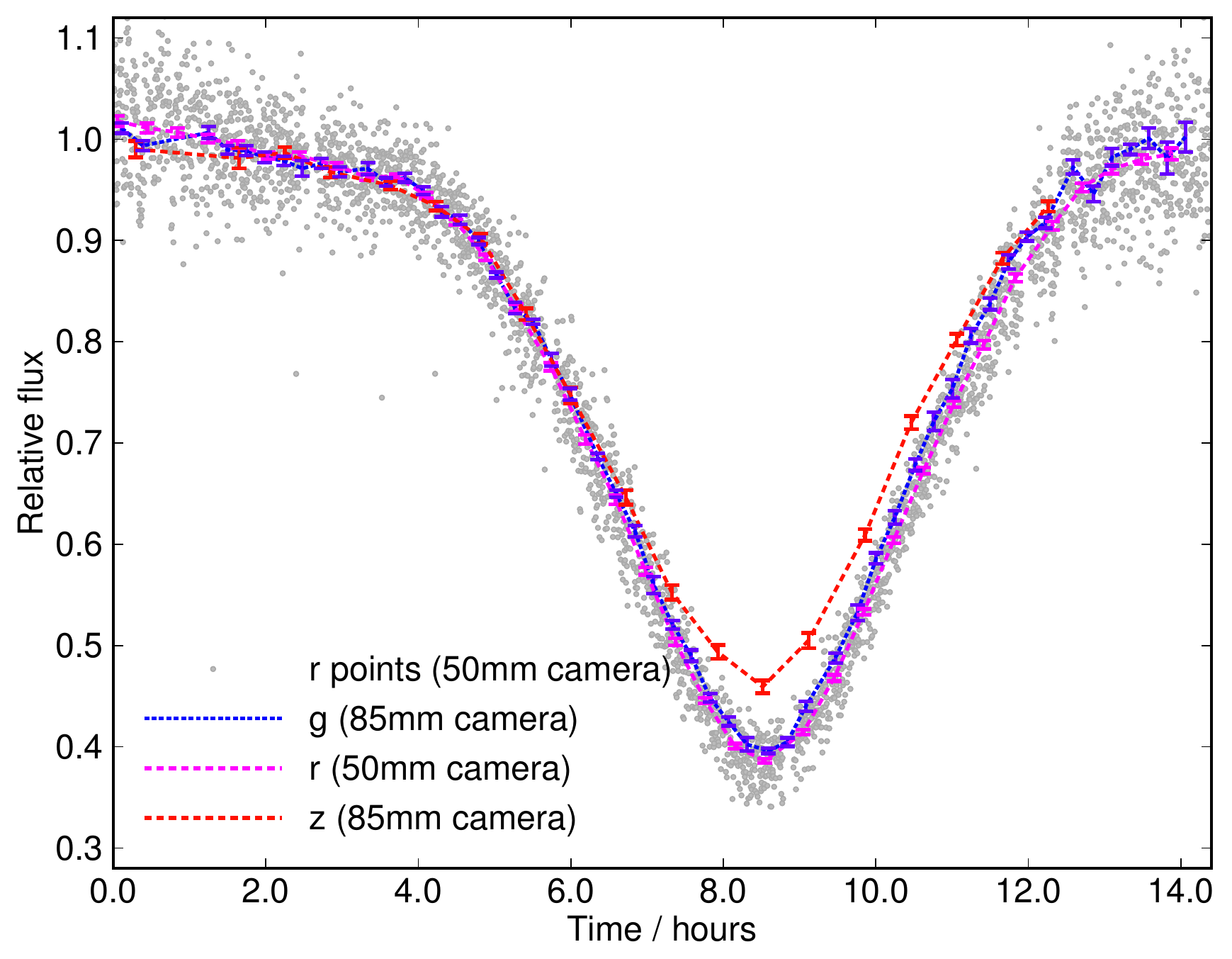}
   }
   \caption{The 18 February 2012 primary eclipse of W UMi, followed over a period of 15 hours by both camera systems. The small grey points are individual 10s exposures from the 50mm camera in the r filter; the lines and points with error bars are binned photometry from each camera, with error bars estimated from the RMS scatter of points within the bin. The red points are near-simultaneous binned photometry taken by the 85mm camera in the z filter, binned over 2$\times$ more points because of the decreased sensitivity of the camera in the z filter. The increase in photometric scatter towards later times is due to the beginning of twilight; the z filter is more affected by this and cuts out earlier. The photometric precision of the binned points is 2-5 millimagnitudes, depending on the filter and camera.
   }

   \label{fig:eclipse}
\end{figure}

Because our targets are bright ($m_V<\sim10$), essentially all the high-amplitude variables in the field are already known.  In an initial visual characterization of each light curve with evidence of a large amount of variability, we re-detected the Delta-Scuti star V377 Cep, and detected eclipses for the beta Lyr eclipsing binaries EG Cep and AZ Cam, the W UMa eclipsing binaries FN Cam and RZ UMi, and the Algol-type Binaries W UMi, TY UMi, AY Cam and SV Cam. In Figure \ref{fig:eclipse} we show a two-camera, three-colour, 15-hour light-curve of the primary eclipse of W UMi, a $\rm m_V = 8.5$ Algol-type eclipsing binary (e.g. \citealt{Sahade1945}). Three primary and secondary eclipses of this 1.71-day-period system were captured during our week-long observations.

\subsection{Stetson-J blind detection of variable stars}

\begin{figure}
  \resizebox{1.025\columnwidth}{!}
   {
	\includegraphics{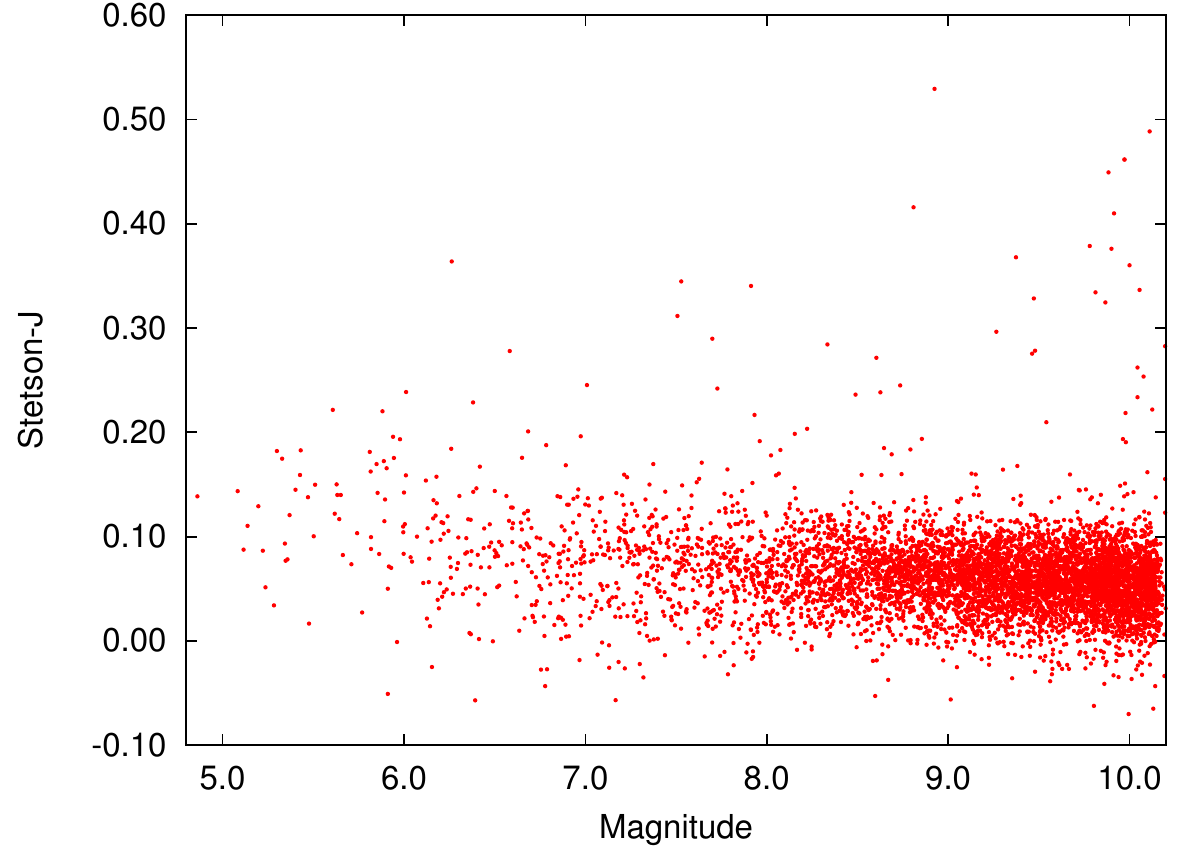}
   }

   \caption{Stetson-J variability index for unbinned light curves of the 5,283 stars in the 50mm-camera exoplanet-search sample. The Stetson-J index for the binned data is similar but with a different normalization.}
   \label{fig:stetson}
\end{figure}

We tested the ability to perform an automated stellar variability and transit search with the cameras, using known bright variable objects in our field to evaluate the detection efficiency. The General Catalogue of Variable Stars (GCVS; \citealt{Samus2009}) lists 119 variable stars within our field of view and magnitude limits ($\rm5.0<m_V<10.0$). Most of these objects are rarely outbursting stars or long-period variables which are unlikely to have produced detectable variability during our observations. However, 28 stars have $<$$20$ day photometric periods (and so a reasonable chance of undergoing measurable variability within our observations), and have light curves which were flagged as clean of instrumental effects by our pipeline. All the known variables in the list have amplitudes which can be easily detected by our cameras.

We based the variability search on the Stetson-J statistic \citep{Stetson1996}, which compares pairs of observations to search for variability which is correlated between datapoints. Compared to a simple RMS-variability selection we found that Stetson-J is much less sensitive to the main source of false-positive variability in our dataset: few-percent-level blending between closely separated sources, which affects approximately 2\% of the stars in the field. Most of the stars in the field move by a large fraction of a pixel each frame, and thus the blending between closely-separated stars can vary by large amounts ($>10\%$) frame-to-frame. The resulting mostly uncorrelated blending from frame-to-frame allows the Stetson-J statistic to exclude those sources as instrumentally rather than astrophysically variable.

To detect a source as variable we set a minimum Stetson-J statistic of 0.14, a value selected to exclude 98\% of the sources in our field (Figure \ref{fig:stetson}). Seven (25\%) of the GCVS variable stars were detected as variable under this criterion. The light curves of the remaining GCVS objects were clean of obvious instrumental effects, but did not contain eclipses or other obvious astrophysical variability, suggesting the pipeline is efficiently and automatically detecting the objects that underwent eclipses or other variability during the observation period evaluated. 

A visual inspection of the light curves of the other 119 high-variability sources revealed eight objects with signs of pulsations or other sinusoidal variability at the $<$5\% level; five objects with possible single eclipses with 5-20\% depths (suggesting long-period systems); 43 sources with blending effects from nearby stars; and 63 objects with long-timescale variability that we cannot usefully classify as instrumental or astrophysical in the current dataset. The planned full-winter Arctic datasets will allow confirmation of these possible longer-period variable stars and enable a full evaluation of stellar variability in this field on timescales up to several months. 

\subsection{Transit search}

We searched for individual planetary transits using the binned high-precision light curves shown in Figure \ref{fig:full_phot}. We calculated the expected Stetson-J transit signal by injecting simulated transit events (1\%, 2-hour-duration dips with ingress and egress periods) into our binned light curves. We found that individual transit events are securely detectable using the Stetson-J statistic for the brightest 1/3 of the stars in the dataset ($\rm m_V\lsim8.5$). Typical 1\%-depth transits produce Stetson-J values of 0.3 in the binned light curves of these bright objects, excluding the 97.5\% of objects in the field which have lower values. A visual search of the 38 light curves with Stetson-J $>$ 0.3 in the binned light curves revealed no likely transit-like events. The statistically-significant detection of transit signals around the fainter stars will be feasible with a longer dataset which allows the use of more sophisticated phased matched-filter-type transit-search algorithms that obtain much improved detection significances (e.g. \citealt{Kovacs2002, Tingley2003}).

\section{Discussion \& conclusions}
\label{sec:disc}

We have demonstrated the operation of two wide-field camera systems in the High Canadian Arctic, achieving the image quality and photometric precision necessary for the detection of transiting exoplanets around bright stars. Each camera's dataset contains light curves for $\approx$70,000 stars, with $\approx$5,000 in each camera having sufficient SNR and being sufficiently uncrowded to search for transiting exoplanets. 

In a dataset from a full arctic winter, the probability of detecting at least three transits of planets in month-long orbits around solar-type stars is $\approx$70\% (Figure \ref{fig:detect_effec}), while the geometric transit probability for those planets is a few percent. With few-millimag photometric precision the systems could detect exoplanets as small as Neptune transiting solar-type stars. The \textit{Kepler}-derived few-percent occurrence frequencies for planets in that size and orbital period range \citep{Howard2011} suggest that the AWCam systems, when operated during an entire arctic winter, will be capable of detecting up to five transiting exoplanets around stars brighter than $\rm m_V$=9.5.

\subsection{A concept for a 10,300 square degree Arctic camera}

Although wide field, the cameras described here only cover 3-6\% of the sky continuously accessible from the Ridge Lab in winter months. Larger sky areas can be covered with multiple cameras, but this requires extremely short exposures or a tracking mount to avoid star trails. One such concept under development, the Compound Arctic Telescope System (CATS, \citealt{Law2012arctic}), places 19 cameras similar to the ones described in this paper in a hemisphere which rotates once per day. This arrangement would cover a total of 10,300 square degrees from one rotating mount, providing full coverage of the Polar sky down to $30^{\circ}$ degrees declination. The system is designed to be capable of monitoring 90,000 stars brighter than 10$^{\rm th}$ magnitude, with millimagnitude precision in each few-minute exposure. A further $\sim$$10^6$ stars can be covered with millimagnitude precision with a 30-minute cadence, and stars brighter than $\rm m_V\approx17$ would be monitored with SNR$>$$5$ every few minutes. The camera has only one major moving part to provide the mount rotation, and so can be constructed to be sufficiently reliable for autonomous operations in the Arctic. Coupled to a small spectrograph-equipped telescope, such as an upgraded Dunlap Institute Arctic Telescope \citep{Law2012, Law2012arctic}, the system would provide an integrated, rapid High Arctic survey and follow-up facility.

\acknowledgments 
\section*{Acknowledgements}

We thank Paul Hickson and James Graham for very useful discussions. It is our pleasure to acknowledge the arctic expertise of Pierre Fogal and James Drummond from the Canadian Network for the Detection of Atmospheric Change, and for their help during operations at the Polar Environment Atmospheric Laboratory. We are also indebted to Environment Canada and the staff of the Eureka weatherstation for their hospitality and support of our observing run. N.M.L. and S.S. are Dunlap Fellows at the Dunlap Institute for Astronomy \& Astrophysics, University of Toronto. The Dunlap Institute is funded through an endowment established by the David Dunlap family and the University of Toronto. This project was partially supported by funds from the Natural Sciences and Engineering Research Council of Canada, and the National Research Council of Canada. The research made use of tools provided by Astrometry.net; NASA's Astrophysics Data System Bibliographic Services; and the SINDBAD database operated at CDS, Strasbourg, France.

\bibliographystyle{apj}
\bibliography{refs}

\begin{thebibliography}{52}
\expandafter\ifx\csname natexlab\endcsname\relax\def\natexlab#1{#1}\fi

\bibitem[{{Agol} {et~al.}(2010){Agol}, {Cowan}, {Knutson}, {Deming}, {Steffen},
  {Henry}, \& {Charbonneau}}]{Agol2010}
{Agol}, E., {Cowan}, N.~B., {Knutson}, H.~A., {et~al.} 2010, \apj, 721, 1861

\bibitem[{{Bakos} {et~al.}(2004){Bakos}, {Noyes}, {Kov{\'a}cs}, {Stanek},
  {Sasselov}, \& {Domsa}}]{Bakos2004}
{Bakos}, G., {Noyes}, R.~W., {Kov{\'a}cs}, G., {et~al.} 2004, \pasp, 116, 266

\bibitem[{{Bakos} {et~al.}(2007){Bakos}, {Kov{\'a}cs}, {Torres}, {Fischer},
  {Latham}, {Noyes}, {Sasselov}, {Mazeh}, {Shporer}, {Butler}, {Stefanik},
  {Fern{\'a}ndez}, {Sozzetti}, {P{\'a}l}, {Johnson}, {Marcy}, {Winn}, {Sip{\H
  o}cz}, {L{\'a}z{\'a}r}, {Papp}, \& {S{\'a}ri}}]{Bakos2007}
{Bakos}, G.~{\'A}., {Kov{\'a}cs}, G., {Torres}, G., {et~al.} 2007, \apj, 670,
  826

\bibitem[{{Benn} \& {Ellison}(1998)}]{Benn1998}
{Benn}, C.~R., \& {Ellison}, S.~L. 1998, New Astronomy Reviews, 42, 503

\bibitem[{{Bertin}(2006)}]{Bertin2006}
{Bertin}, E. 2006, in Astronomical Society of the Pacific Conference Series,
  Vol. 351, Astronomical Data Analysis Software and Systems XV, ed.
  {C.~Gabriel, C.~Arviset, D.~Ponz, \& S.~Enrique}, 112

\bibitem[{{Bertin} \& {Arnouts}(1996)}]{Bertin1996}
{Bertin}, E., \& {Arnouts}, S. 1996, \aaps, 117, 393

\bibitem[{{Bouchy} {et~al.}(2005){Bouchy}, {Udry}, {Mayor}, {Moutou}, {Pont},
  {Iribarne}, {da Silva}, {Ilovaisky}, {Queloz}, {Santos}, {S{\'e}gransan}, \&
  {Zucker}}]{Bouchy2005}
{Bouchy}, F., {Udry}, S., {Mayor}, M., {et~al.} 2005, \aap, 444, L15

\bibitem[{{Brown}(1966)}]{Brown1966}
{Brown}, D.~C. 1966, Photometric Engineering, 32, 444

\bibitem[{{Brown}(2003)}]{Brown2003}
{Brown}, T.~M. 2003, \apjl, 593, L125

\bibitem[{{Charbonneau} {et~al.}(2000){Charbonneau}, {Brown}, {Latham}, \&
  {Mayor}}]{Charbonneau2000}
{Charbonneau}, D., {Brown}, T.~M., {Latham}, D.~W., \& {Mayor}, M. 2000, \apjl,
  529, L45

\bibitem[{{Christian} {et~al.}(2006){Christian}, {Pollacco}, {Skillen},
  {Street}, {Keenan}, {Clarkson}, {Collier Cameron}, {Kane}, {Lister}, {West},
  {Enoch}, {Evans}, {Fitzsimmons}, {Haswell}, {Hellier}, {Hodgkin}, {Horne},
  {Irwin}, {Norton}, {Osborne}, {Ryans}, {Wheatley}, \&
  {Wilson}}]{Christian2006}
{Christian}, D.~J., {Pollacco}, D.~L., {Skillen}, I., {et~al.} 2006, \mnras,
  372, 1117

\bibitem[{{Collier Cameron} {et~al.}(2010){Collier Cameron}, {Guenther},
  {Smalley}, {McDonald}, {Hebb}, {Andersen}, {Augusteijn}, {Barros}, {Brown},
  {Cochran}, {Endl}, {Fossey}, {Hartmann}, {Maxted}, {Pollacco}, {Skillen},
  {Telting}, {Waldmann}, \& {West}}]{Collier2010}
{Collier Cameron}, A., {Guenther}, E., {Smalley}, B., {et~al.} 2010, \mnras,
  407, 507

\bibitem[{{Conrady}(1919)}]{Conrady1919}
{Conrady}, A.~E. 1919, \mnras, 79, 384

\bibitem[{{Crouzet} {et~al.}(2010){Crouzet}, {Guillot}, {Agabi}, {Rivet},
  {Bondoux}, {Challita}, {Fante{\"i}-Caujolle}, {Fressin}, {M{\'e}karnia},
  {Schmider}, {Valbousquet}, {Blazit}, {Bonhomme}, {Abe}, {Daban}, {Gouvret},
  {Fruth}, {Rauer}, {Erikson}, {Barbieri}, {Aigrain}, \& {Pont}}]{Crouzet2010}
{Crouzet}, N., {Guillot}, T., {Agabi}, A., {et~al.} 2010, \aap, 511, A36

\bibitem[{{Daban} {et~al.}(2010){Daban}, {Gouvret}, {Guillot}, {Agabi},
  {Crouzet}, {Rivet}, {Mekarnia}, {Abe}, {Bondoux}, {Fante{\"i}-Caujolle},
  {Fressin}, {Schmider}, {Valbousquet}, {Blanc}, {Le van Suu}, {Rauer},
  {Erikson}, {Pont}, \& {Aigrain}}]{Daban2010}
{Daban}, J.-B., {Gouvret}, C., {Guillot}, T., {et~al.} 2010, in Society of
  Photo-Optical Instrumentation Engineers (SPIE) Conference Series, Vol. 7733,
  Society of Photo-Optical Instrumentation Engineers (SPIE) Conference Series

\bibitem[{{Demory} {et~al.}(2011){Demory}, {Gillon}, {Deming}, {Valencia},
  {Seager}, {Benneke}, {Lovis}, {Cubillos}, {Harrington}, {Stevenson}, {Mayor},
  {Pepe}, {Queloz}, {S{\'e}gransan}, \& {Udry}}]{Demory2011}
{Demory}, B.-O., {Gillon}, M., {Deming}, D., {et~al.} 2011, \aap, 533, A114

\bibitem[{{di Rico} {et~al.}(2010){di Rico}, {Ragni}, {Dolci}, {Straniero},
  {Valentini}, {Valentini}, {di Cianno}, {Giuliani}, {Bonoli}, {Bortoletto},
  {D'Alessandro}, {Magrin}, {Corcione}, {Riva}, {Abia}, {Mancini}, {Busso}, \&
  {Tosti}}]{Rico2010}
{di Rico}, G., {Ragni}, M., {Dolci}, M., {et~al.} 2010, in Society of
  Photo-Optical Instrumentation Engineers (SPIE) Conference Series, Vol. 7737,
  Society of Photo-Optical Instrumentation Engineers (SPIE) Conference Series

\bibitem[{{Dravins} {et~al.}(1998){Dravins}, {Lindegren}, {Mezey}, \&
  {Young}}]{Dravins1998}
{Dravins}, D., {Lindegren}, L., {Mezey}, E., \& {Young}, A.~T. 1998, \pasp,
  110, 610

\bibitem[{{Henry} {et~al.}(2000){Henry}, {Marcy}, {Butler}, \&
  {Vogt}}]{Henry2000}
{Henry}, G.~W., {Marcy}, G.~W., {Butler}, R.~P., \& {Vogt}, S.~S. 2000, \apjl,
  529, L41

\bibitem[{{Hickson} {et~al.}(2010){Hickson}, {Carlberg}, {Gagne}, {Pfrommer},
  {Racine}, {Sch{\"o}ck}, {Steinbring}, \& {Travouillon}}]{Hickson2010}
{Hickson}, P., {Carlberg}, R., {Gagne}, R., {et~al.} 2010, in Society of
  Photo-Optical Instrumentation Engineers (SPIE) Conference Series, Vol. 7733,
  Society of Photo-Optical Instrumentation Engineers (SPIE) Conference Series

\bibitem[{{Howard} {et~al.}(2011){Howard}, {Marcy}, {Bryson}, {Jenkins},
  {Rowe}, {Batalha}, {Borucki}, {Koch}, {Dunham}, {Gautier}, {Van Cleve},
  {Cochran}, {Latham}, {Lissauer}, {Torres}, {Brown}, {Gilliland}, {Buchhave},
  {Caldwell}, {Christensen-Dalsgaard}, {Ciardi}, {Fressin}, {Haas}, {Howell},
  {Kjeldsen}, {Seager}, {Rogers}, {Sasselov}, {Steffen}, {Basri},
  {Charbonneau}, {Christiansen}, {Clarke}, {Dupree}, {Fabrycky}, {Fischer},
  {Ford}, {Fortney}, {Tarter}, {Girouard}, {Holman}, {Johnson}, {Klaus},
  {Machalek}, {Moorhead}, {Morehead}, {Ragozzine}, {Tenenbaum}, {Twicken},
  {Quinn}, {Isaacson}, {Shporer}, {Lucas}, {Walkowicz}, {Welsh}, {Boss},
  {Devore}, {Gould}, {Smith}, {Morris}, {Prsa}, \& {Morton}}]{Howard2011}
{Howard}, A.~W., {Marcy}, G.~W., {Bryson}, S.~T., {et~al.} 2011, ArXiv e-prints

\bibitem[{{Howell} {et~al.}(2012){Howell}, {Rowe}, {Bryson}, {Quinn}, {Marcy},
  {Isaacson}, {Ciardi}, {Chaplin}, {Metcalfe}, {Monteiro}, {Appourchaux},
  {Basu}, {Creevey}, {Gilliland}, {Quirion}, {Stello}, {Kjeldsen},
  {Christensen-Dalsgaard}, {Elsworth}, {Garc{\'{\i}}a}, {Houdek}, {Karoff},
  {Molenda-{\.Z}akowicz}, {Thompson}, {Verner}, {Torres}, {Fressin}, {Crepp},
  {Adams}, {Dupree}, {Sasselov}, {Dressing}, {Borucki}, {Koch}, {Lissauer},
  {Latham}, {Buchhave}, {Gautier}, {Everett}, {Horch}, {Batalha}, {Dunham},
  {Szkody}, {Silva}, {Mighell}, {Holberg}, {Ballot}, {Bedding}, {Bruntt},
  {Campante}, {Handberg}, {Hekker}, {Huber}, {Mathur}, {Mosser}, {R{\'e}gulo},
  {White}, {Christiansen}, {Middour}, {Haas}, {Hall}, {Jenkins}, {McCaulif},
  {Fanelli}, {Kulesa}, {McCarthy}, \& {Henze}}]{Howell2012}
{Howell}, S.~B., {Rowe}, J.~F., {Bryson}, S.~T., {et~al.} 2012, \apj, 746, 123

\bibitem[{{Kov{\'a}cs} {et~al.}(2002){Kov{\'a}cs}, {Zucker}, \&
  {Mazeh}}]{Kovacs2002}
{Kov{\'a}cs}, G., {Zucker}, S., \& {Mazeh}, T. 2002, \aap, 391, 369

\bibitem[{{Lang} {et~al.}(2010){Lang}, {Hogg}, {Mierle}, {Blanton}, \&
  {Roweis}}]{Lang2010}
{Lang}, D., {Hogg}, D.~W., {Mierle}, K., {Blanton}, M., \& {Roweis}, S. 2010,
  \aj, 139, 1782

\bibitem[{{Law} {et~al.}(2009){Law}, {Kulkarni}, {Dekany}, {Ofek}, {Quimby},
  {Nugent}, {Surace}, {Grillmair}, {Bloom}, {Kasliwal}, {Bildsten}, {Brown},
  {Cenko}, {Ciardi}, {Croner}, {Djorgovski}, {van Eyken}, {Filippenko}, {Fox},
  {Gal-Yam}, {Hale}, {Hamam}, {Helou}, {Henning}, {Howell}, {Jacobsen},
  {Laher}, {Mattingly}, {McKenna}, {Pickles}, {Poznanski}, {Rahmer}, {Rau},
  {Rosing}, {Shara}, {Smith}, {Starr}, {Sullivan}, {Velur}, {Walters}, \&
  {Zolkower}}]{Law09}
{Law}, N.~M., {Kulkarni}, S.~R., {Dekany}, R.~G., {et~al.} 2009, \pasp, 121,
  1395

\bibitem[{{Law} {et~al.}(2011){Law}, {Kraus}, {Street}, {Fulton},
  {Hillenbrand}, {Shporer}, {Lister}, {Baranec}, {Bloom}, {Bui}, {Burse},
  {Cenko}, {Das}, {Davis}, {Dekany}, {Filippenko}, {Kasliwal}, {Kulkarni},
  {Nugent}, {Ofek}, {Poznanski}, {Quimby}, {Ramaprakash}, {Riddle},
  {Silverman}, {Sivanandam}, \& {Tendulkar}}]{Law2012}
{Law}, N.~M., {Kraus}, A.~L., {Street}, R., {et~al.} 2011, ArXiv e-prints

\bibitem[{{Law} {et~al.}(2012){Law}, {Sivanandam}, {Murowinski}, {Carlberg},
  {Ngan}, {Salbi}, {Ahmadi}, {Steinbring}, {Halman}, \&
  {Graham}}]{Law2012arctic}
{Law}, N.~M., {Sivanandam}, S., {Murowinski}, R., {et~al.} 2012, ArXiv e-prints

\bibitem[{{Majeau} {et~al.}(2012){Majeau}, {Agol}, \& {Cowan}}]{Majeau2012}
{Majeau}, C., {Agol}, E., \& {Cowan}, N.~B. 2012, \apjl, 747, L20

\bibitem[{{Moore} {et~al.}(2006){Moore}, {Aristidi}, {Ashley}, {Busso},
  {Candidi}, {Everett}, {Kenyon}, {Lawrence}, {Luong-Van}, {Phillips}, {Le
  Roux}, {Ragazzoni}, {Salinari}, {Storey}, {Taylor}, {Tosti}, \&
  {Travouillon}}]{Moore2006}
{Moore}, A., {Aristidi}, E., {Ashley}, M., {et~al.} 2006, in Society of
  Photo-Optical Instrumentation Engineers (SPIE) Conference Series, Vol. 6267,
  Society of Photo-Optical Instrumentation Engineers (SPIE) Conference Series

\bibitem[{{Moore} {et~al.}(2008){Moore}, {Allen}, {Aristidi}, {Ashley},
  {Bedding}, {Beichman}, {Briguglio}, {Busso}, {Candidi}, {Ciardi}, {Cui},
  {Cutispoto}, {Distefano}, {Espy}, {Everett}, {Feng}, {Hu}, {Jiang}, {Kenyon},
  {Kulesa}, {Lawrence}, {Le Roux}, {Leslie}, {Li}, {Luong-Van}, {Phillips},
  {Qin}, {Ragazzoni}, {Riddle}, {Sabbatini}, {Salinari}, {Saunders}, {Shang},
  {Stello}, {Storey}, {Sun}, {Suntzeff}, {Taylor}, {Tosti}, {Tothill},
  {Travouillon}, {Van Belle}, {Von Braun}, {Wang}, {Yan}, {Yang}, {Yuan},
  {Zhu}, \& {Zhou}}]{Moore2008}
{Moore}, A., {Allen}, G., {Aristidi}, E., {et~al.} 2008, in Society of
  Photo-Optical Instrumentation Engineers (SPIE) Conference Series, Vol. 7012,
  Society of Photo-Optical Instrumentation Engineers (SPIE) Conference Series

\bibitem[{{Moore} {et~al.}(2010){Moore}, {Ahmed}, {Ashley}, {Barreto}, {Cui},
  {Delacroix}, {Feng}, {Gong}, {Lawrence}, {Luong-van}, {Martin}, {Riddle},
  {Rowley}, {Shang}, {Storey}, {Tothill}, {Travouillon}, {Wang}, {Yang},
  {Yang}, {Zhou}, \& {Zhu}}]{Moore2010}
{Moore}, A.~M., {Ahmed}, S., {Ashley}, M.~C.~B., {et~al.} 2010, in Society of
  Photo-Optical Instrumentation Engineers (SPIE) Conference Series, Vol. 7733,
  Society of Photo-Optical Instrumentation Engineers (SPIE) Conference Series

\bibitem[{{P{\'a}l} {et~al.}(2010){P{\'a}l}, {Bakos}, {Torres}, {Noyes},
  {Fischer}, {Johnson}, {Henry}, {Butler}, {Marcy}, {Howard}, {Sip{\H o}cz},
  {Latham}, \& {Esquerdo}}]{Pal2010}
{P{\'a}l}, A., {Bakos}, G.~{\'A}., {Torres}, G., {et~al.} 2010, \mnras, 401,
  2665

\bibitem[{{Pepper} {et~al.}(2012){Pepper}, {Kuhn}, {Siverd}, {James}, \&
  {Stassun}}]{Pepper2012}
{Pepper}, J., {Kuhn}, R.~B., {Siverd}, R., {James}, D., \& {Stassun}, K. 2012,
  \pasp, 124, 230

\bibitem[{{Pepper} {et~al.}(2007){Pepper}, {Pogge}, {DePoy}, {Marshall},
  {Stanek}, {Stutz}, {Poindexter}, {Siverd}, {O'Brien}, {Trueblood}, \&
  {Trueblood}}]{Pepper2007}
{Pepper}, J., {Pogge}, R.~W., {DePoy}, D.~L., {et~al.} 2007, \pasp, 119, 923

\bibitem[{{Pollacco} {et~al.}(2006){Pollacco}, {Skillen}, {Collier Cameron},
  {Christian}, {Hellier}, {Irwin}, {Lister}, {Street}, {West}, {Anderson},
  {Clarkson}, {Deeg}, {Enoch}, {Evans}, {Fitzsimmons}, {Haswell}, {Hodgkin},
  {Horne}, {Kane}, {Keenan}, {Maxted}, {Norton}, {Osborne}, {Parley}, {Ryans},
  {Smalley}, {Wheatley}, \& {Wilson}}]{Pollacco2006}
{Pollacco}, D.~L., {Skillen}, I., {Collier Cameron}, A., {et~al.} 2006, \pasp,
  118, 1407

\bibitem[{{Pont} \& {Bouchy}(2005)}]{Pont2005}
{Pont}, F., \& {Bouchy}, F. 2005, in EAS Publications Series, Vol.~14, EAS
  Publications Series, ed. {M.~Giard, F.~Casoli, \& F.~Paletou}, 155--160

\bibitem[{{Sahade}(1945)}]{Sahade1945}
{Sahade}, J. 1945, \apj, 102, 470

\bibitem[{{Samus} {et~al.}(2009){Samus}, {Durlevich}, \& {et al.}}]{Samus2009}
{Samus}, N.~N., {Durlevich}, O.~V., \& {et al.} 2009, VizieR Online Data
  Catalog, 1, 2025

\bibitem[{{Sato} {et~al.}(2005){Sato}, {Fischer}, {Henry}, {Laughlin},
  {Butler}, {Marcy}, {Vogt}, {Bodenheimer}, {Ida}, {Toyota}, {Wolf}, {Valenti},
  {Boyd}, {Johnson}, {Wright}, {Ammons}, {Robinson}, {Strader}, {McCarthy},
  {Tah}, \& {Minniti}}]{Sato2005}
{Sato}, B., {Fischer}, D.~A., {Henry}, G.~W., {et~al.} 2005, \apj, 633, 465

\bibitem[{{Smith} {et~al.}(2006){Smith}, {Collier Cameron}, {Christian},
  {Clarkson}, {Enoch}, {Evans}, {Haswell}, {Hellier}, {Horne}, {Irwin}, {Kane},
  {Lister}, {Norton}, {Parley}, {Pollacco}, {Ryans}, {Skillen}, {Street},
  {Triaud}, {West}, {Wheatley}, \& {Wilson}}]{Smith2006}
{Smith}, A.~M.~S., {Collier Cameron}, A., {Christian}, D.~J., {et~al.} 2006,
  \mnras, 373, 1151

\bibitem[{{Steinbring} {et~al.}(2012){Steinbring}, {Ward}, \&
  {Drummond}}]{Steinbring2012}
{Steinbring}, E., {Ward}, W., \& {Drummond}, J.~R. 2012, \pasp, 124, 185

\bibitem[{{Steinbring} {et~al.}(2010){Steinbring}, {Carlberg}, {Croll},
  {Fahlman}, {Hickson}, {Ivanescu}, {Leckie}, {Pfrommer}, \&
  {Schoeck}}]{Steinbring2010}
{Steinbring}, E., {Carlberg}, R., {Croll}, B., {et~al.} 2010, \pasp, 122, 1092

\bibitem[{{Stetson}(1996)}]{Stetson1996}
{Stetson}, P.~B. 1996, \pasp, 108, 851

\bibitem[{{Stevenson} {et~al.}(2011){Stevenson}, {Harrington}, {Fortney},
  {Loredo}, {Hardy}, {Nymeyer}, {Bowman}, {Cubillos}, {Bowman}, \&
  {Hardin}}]{Stevenson2011}
{Stevenson}, K.~B., {Harrington}, J., {Fortney}, J., {et~al.} 2011, ArXiv
  e-prints

\bibitem[{{Tamuz} {et~al.}(2005){Tamuz}, {Mazeh}, \& {Zucker}}]{Tamuz2005}
{Tamuz}, O., {Mazeh}, T., \& {Zucker}, S. 2005, \mnras, 356, 1466

\bibitem[{{Tingley}(2003)}]{Tingley2003}
{Tingley}, B. 2003, \aap, 408, L5

\bibitem[{{von Braun} {et~al.}(2011){von Braun}, {Boyajian}, {ten Brummelaar},
  {Kane}, {van Belle}, {Ciardi}, {Raymond}, {L{\'o}pez-Morales}, {McAlister},
  {Schaefer}, {Ridgway}, {Sturmann}, {Sturmann}, {White}, {Turner},
  {Farrington}, \& {Goldfinger}}]{vonBraun2011}
{von Braun}, K., {Boyajian}, T.~S., {ten Brummelaar}, T.~A., {et~al.} 2011,
  \apj, 740, 49

\bibitem[{{Wang} {et~al.}(2011){Wang}, {Macri}, {Krisciunas}, {Wang}, {Ashley},
  {Cui}, {Feng}, {Gong}, {Lawrence}, {Liu}, {Luong-Van}, {Pennypacker},
  {Shang}, {Storey}, {Yang}, {Yang}, {Yuan}, {York}, {Zhou}, {Zhu}, \&
  {Zhu}}]{Wang2011}
{Wang}, L., {Macri}, L.~M., {Krisciunas}, K., {et~al.} 2011, \aj, 142, 155

\bibitem[{{Winn} {et~al.}(2011){Winn}, {Matthews}, {Dawson}, {Fabrycky},
  {Holman}, {Kallinger}, {Kuschnig}, {Sasselov}, {Dragomir}, {Guenther},
  {Moffat}, {Rowe}, {Rucinski}, \& {Weiss}}]{Winn2011}
{Winn}, J.~N., {Matthews}, J.~M., {Dawson}, R.~I., {et~al.} 2011, \apjl, 737,
  L18

\bibitem[{{Yanny} {et~al.}(2009){Yanny}, {Rockosi}, {Newberg}, {Knapp},
  {Adelman-McCarthy}, {Alcorn}, {Allam}, {Allende Prieto}, {An}, {Anderson},
  {Anderson}, {Bailer-Jones}, {Bastian}, {Beers}, {Bell}, {Belokurov},
  {Bizyaev}, {Blythe}, {Bochanski}, {Boroski}, {Brinchmann}, {Brinkmann},
  {Brewington}, {Carey}, {Cudworth}, {Evans}, {Evans}, {Gates}, {G{\"a}nsicke},
  {Gillespie}, {Gilmore}, {Nebot Gomez-Moran}, {Grebel}, {Greenwell}, {Gunn},
  {Jordan}, {Jordan}, {Harding}, {Harris}, {Hendry}, {Holder}, {Ivans},
  {Ivezi{\v c}}, {Jester}, {Johnson}, {Kent}, {Kleinman}, {Kniazev},
  {Krzesinski}, {Kron}, {Kuropatkin}, {Lebedeva}, {Lee}, {French Leger},
  {L{\'e}pine}, {Levine}, {Lin}, {Long}, {Loomis}, {Lupton}, {Malanushenko},
  {Malanushenko}, {Margon}, {Martinez-Delgado}, {McGehee}, {Monet}, {Morrison},
  {Munn}, {Neilsen}, {Nitta}, {Norris}, {Oravetz}, {Owen}, {Padmanabhan},
  {Pan}, {Peterson}, {Pier}, {Platson}, {Re Fiorentin}, {Richards}, {Rix},
  {Schlegel}, {Schneider}, {Schreiber}, {Schwope}, {Sibley}, {Simmons},
  {Snedden}, {Allyn Smith}, {Stark}, {Stauffer}, {Steinmetz}, {Stoughton},
  {SubbaRao}, {Szalay}, {Szkody}, {Thakar}, {Thirupathi}, {Tucker}, {Uomoto},
  {Vanden Berk}, {Vidrih}, {Wadadekar}, {Watters}, {Wilhelm}, {Wyse}, {Yarger},
  \& {Zucker}}]{Yanny2009}
{Yanny}, B., {Rockosi}, C., {Newberg}, H.~J., {et~al.} 2009, \aj, 137, 4377

\bibitem[{{Young}(1967)}]{Young1967}
{Young}, A.~T. 1967, \aj, 72, 747

\bibitem[{{Zacharias} {et~al.}(2010){Zacharias}, {Finch}, {Girard}, {Hambly},
  {Wycoff}, {Zacharias}, {Castillo}, {Corbin}, {DiVittorio}, {Dutta}, {Gaume},
  {Gauss}, {Germain}, {Hall}, {Hartkopf}, {Hsu}, {Holdenried}, {Makarov},
  {Martinez}, {Mason}, {Monet}, {Rafferty}, {Rhodes}, {Siemers}, {Smith},
  {Tilleman}, {Urban}, {Wieder}, {Winter}, \& {Young}}]{Zacharias2010}
{Zacharias}, N., {Finch}, C., {Girard}, T., {et~al.} 2010, \aj, 139, 2184

\end{thebibliography}

\label{lastpage}

\end{document}